\documentclass[a4paper]{jpconf}
\usepackage{graphicx}
\begin{document}
\title{ Normal and lateral Casimir force: Advances and prospects}

\author{G. L. Klimchitskaya}

\address{Department of Physics, North-West Technical University,
Millionnaya Street 5, St.Petersburg,
191065, Russia {\protect \\}
and {\protect \\}Institute for Theoretical
Physics, Leipzig University, Postfach 100920,
D-04009, Leipzig, Germany}

\ead{galina.klimchitskaya@itp.uni-leipzig.de}

\begin{abstract}
We discuss recent experimental and theoretical results on the Casimir force
between real material bodies made of different materials. Special attention is paid
to calculations of the normal Casimir force acting perpendicular to the
surface with the help of the Lifshitz theory taking into account the role
of free charge carriers. Theoretical results for the thermal Casimir
force acting between metallic, dielectric and semiconductor materials
are presented and compared with available experimental data.
Main attention is concentrated on the possibility to control the
magnitude and sign of the Casimir force for applications in nanotechnology.
In this respect we consider experiments on the optical modulation of the
Casimir force between metal and semiconductor test bodies with laser
light. Another option is the use of ferromagnetic materials,
specifically, ferromagnetic dielectrics. Under some conditions this
allows to get Casimir repulsion. The lateral Casimir force acting
between sinusoidally corrugated surfaces can be considered as some
kind of noncontact friction caused by zero-point oscillations of the
electromagnetic field. Recent experiments and computations using the exact
theory have demonstrated the role of diffraction-type effects in this
phenomenon and the possibility to get asymmetric force profiles.
Conclusion is made that the Casimir force may play important role in the
operation of different devices on the nanoscale.
\end{abstract}

\section{Introduction}

Both the van der Waals and Casimir forces are quantum phenomena
originating from the interaction of electromagnetic fluctuations
with matter. One difference between them is that the van der
Waals force is nonrelativistic, whereas the Casimir force
takes full account of the relativistic retardation effects.
Thus, in fact there exists just one force of a fluctuation nature.
At shortest sepatations up to a few nanometers between the
interacting bodies, where retardation is negligibly small, it is
referred to as the {\it van der Waals} force. At larger separations it
is called the {\it  Casimir} force \cite{1}.

The Casimir force becomes significantly large if measured at
separations below one micrometer. At separations below 100\,nm
it competes with characteristic electric forces acting
between the constituting elements of microelectromechanical and
nanoelectromechanical devices. The attractive Casimir force leads
to pull-in and stiction \cite{2} which may result in the collapse
of a microdevice. That is why the Casimir force has attracted much
recent attention in nanotechnology. On the one hand, it can be
used as an actuating force \cite{3} in place of electric forces.
On the other hand, it plays the role of some kind friction in
micrometer scales which cannot be diminished by means of lubrication.
In this respect, the possibility of repulsive Casimir force
attracts much public attention as the mean to cancel this
friction.

It is commonly known that the unified description of both
the van der Waals and Casimir forces is given by the Lifshitz
theory \cite{4,5}. During more than fourty years after this
theory was created it was lacking detailed experimental
confirmation due to the absence of sufficiently precise
measurements. During the last decade, however, many experiments
measuring the Casimir force have been performed (see the recent
review \cite{6}). The comparison between the measurement data of
the most precise experiments and computational results on the
basis of the Lifshitz theory has led to a puzzle.
It turned out that the Lifshitz theory is in agreement with the
data only under a condition that for metallic test bodies the
relaxation properties of conduction electrons are not accounted
for, whereas for dielectrics charge carriers are completely
discarded (controversial discussions concerning these results
are reflected in Ref.~\cite{7}). Modern overview of both
experimental and theoretical aspects of the Casimir force can
be found in the monograph \cite{7a}.

The present paper reviews several recent experimental and
theoretical results in the Casimir effect which are of much
importance for understanding of delicate physical mechanisms
of fluctuation-induced forces and for their technological
application in new generations of micro- and nanoelectromechanical
systems. In Sec.~2 we briefly discuss the standard formalism of
the Lifshitz theory, its generalization for the case of
nonplanar geometries, and modifications aimed to resolve the
above-mentioned puzzle. Section 3 is devoted to the comparison
of theoretical results with the experimental data of three
landmark experiments. These experiments were performed with
metallic \cite{8,9,10,11}, semiconductor \cite{12,13}, and
dielectric \cite{14,15} test bodies. In Sec.~4 the possibilities
to control the magnitude and sign of the Casimir force by
using materials with different electric and magnetic properties
are considered. These include the variation of the density of free
charge carriers with laser light, the use of three-layer systems
and phase transitions, investigation of the Casimir force
between ferromagnetic dielectrics etc. Special attention is
devoted in Sec.~5 to the lateral Casimir force arising between
sinusoidally corrugated surfaces, as predicted theoretically
in Ref.~\cite{16} and experimentally demonstrated in
Refs.~\cite{17,18}. It was shown quite recently \cite{19,20}
that this phenomenon has potential for applications
in nanotechnology. Our conclusions and discussion are
contained in Sec.~6.

\section{Brief formulation of the Lifshitz theory}

The Lifshitz theory \cite{4,5} expresses the free energy and
pressure of the van der Waals and Casimir interaction in the
configuration of two parallel semispaces (thick plates)
separated with a gap of width $a$ at temperature $T$ in the
following form:
\begin{eqnarray}
&&
{\cal F}(a,T)=\frac{k_BT}{2\pi}\sum_{l=0}^{\infty}
{\vphantom{\sum}}^{\prime}\Phi_E(\xi_l),
\label{eq1} \\
&&
P(a,T)=-\frac{\partial{\cal F}(a,T)}{\partial a}=
-\frac{k_BT}{\pi}\sum_{l=0}^{\infty}
{\vphantom{\sum}}^{\prime}\Phi_P(\xi_l).
\nonumber
\end{eqnarray}
\noindent
Here, $k_B$ is the Boltzmann constant,
$\xi_l=2\pi k_BTl/\hbar$ with $l=0,\,1,\,2,\,\ldots$ are
the Matsubara frequencies, and the terms with $l=0$ in
the primed sums are multiplied by 1/2. The functions
$\Phi_E$ and $\Phi_P$ are expressed in terms of reflection
coefficients on the two semispaces $r_{\alpha}^{(1)}$ and
$r_{\alpha}^{(2)}$ for two independent polarizations of the
electromagnetic field, transverse magnetic ($\alpha={\rm TM}$)
and transverse electric ($\alpha={\rm TE}$),
\begin{eqnarray}
&&
\Phi_E(x)=\int_{0}^{\infty}k_{\bot}dk_{\bot}\sum_{\alpha}
\ln\left[1-r_{\alpha}^{(1)}(ix,k_{\bot})
r_{\alpha}^{(2)}(ix,k_{\bot})\,e^{-2aq}\right],
\nonumber \\
&&
\Phi_P(x)=\int_{0}^{\infty}k_{\bot}dk_{\bot}q\sum_{\alpha}
\left[\frac{e^{2aq}}{r_{\alpha}^{(1)}(ix,k_{\bot})
r_{\alpha}^{(2)}(ix,k_{\bot})}-1\right]^{-1},
\label{eq2}
\end{eqnarray}
\noindent
where
\begin{equation}
q\equiv q(ix,k_{\bot})=\sqrt{k_{\bot}^2+x^2/c^2},
\label{eq2a}
\end{equation}
\noindent
and $\mbox{\boldmath$k$}_{\bot}=(k_x,k_y)$ is the projection of a
wave vector on the plane of plates.
The reflection coefficients on homogeneous media described by
the frequency-dependent dielectric permittivity
$\varepsilon^{(n)}(\omega)$ and magnetic permeability
$\mu^{(n)}(\omega)$ ($n=1,\,2$) are given by
\begin{equation}
r_{\rm TM}^{(n)}(ix,k_{\bot})=
\frac{\varepsilon^{(n)}(ix)q-k^{(n)}}{\varepsilon^{(n)}(ix)q+k^{(n)}},
\qquad
r_{\rm TE}^{(n)}(ix,k_{\bot})=
\frac{\mu^{(n)}(ix)q-k^{(n)}}{\mu^{(n)}(ix)q+k^{(n)}},
\label{eq3}
\end{equation}
\noindent
where
\begin{equation}
k^{(n)}\equiv k^{(n)}(ix,k_{\bot})=\sqrt{k_{\bot}^2+
\varepsilon^{(n)}(ix)\mu^{(n)}(ix)\frac{x^2}{c^2}}.
\label{eq4}
\end{equation}

The Lifshitz-type formulas similar to (\ref{eq1}) hold also
for the Casimir-Polder free energy and force acting between an
atom with dynamic electric polarizability $\alpha(\omega)$ and
magnetic susceptibility $\beta(\omega)$ and a wall decribed
by $\varepsilon(\omega)$ and $\mu(\omega)$
\begin{eqnarray}
&&
{\cal F}_A(a,T)=-k_BT\sum_{l=0}^{\infty}
{\vphantom{\sum}}^{\prime}
\int_{0}^{\infty}k_{\bot}dk_{\bot}q_l\Phi_A(\xi_l,k_{\bot}),
\label{eq5} \\
&&
F_A(a,T)=-k_BT\sum_{l=0}^{\infty}
{\vphantom{\sum}}^{\prime}
\int_{0}^{\infty}k_{\bot}dk_{\bot}q_l^2\Phi_A(\xi_l,k_{\bot}),
\nonumber
\end{eqnarray}
\noindent
where $q_l=q(i\xi_l,k_{\bot})$.
The function $\Phi_A(\xi_l,k_{\bot})$ is defined as \cite{21}
\begin{eqnarray}
\Phi_A(x,k_{\bot})&=&e^{-2aq}\left\{
\vphantom{\frac{x^2}{q^2c^2}}
2\left[\alpha(ix)\,r_{\rm TM}(ix,k_{\bot})+
\beta(ix)\,r_{\rm TE}(ix,k_{\bot})\right]\right.
\label{eq6} \\
&-&\left.
\frac{x^2}{q^2c^2}
\left[\alpha(ix)+\beta(ix)\right]\,\left[r_{\rm TM}(ix,k_{\bot})+
r_{\rm TE}(ix,k_{\bot})\right]\right\}.
\nonumber
\end{eqnarray}
\noindent
Due to the presence of only one wall, the upper index on the
reflection coefficients is omitted.

Using the so-called {\it proximity force approximation} (PFA), the
Casimir force acting between a sphere and a plate can be expressed
as
\begin{equation}
F(a,T)=2\pi R{\cal F}(a,T),
\label{eq7}
\end{equation}
\noindent
where the free energy  per unit area
for two parallel semispaces is defined in
Eq.~(\ref{eq1}). Under the condition that $a\ll R$, where $R$ is
the sphere radius and $a$ is the closest separation between the
surfaces of a sphere and a plate,
the relative error of the approximate
Eq.~(\ref{eq7}) is less than $a/R$ \cite{6,7a}. Equation (\ref{eq7})
is widely used for the comparison of the experimental data obtained
from the measurements of the Casimir force between a sphere and
a plate with the predictions of the Lifshitz theory.

The original formulation of the Lifshitz theory for two parallel
semispaces separated with a gap can be simply generalized for an
arbitrary number of planar layers of different thickness made of
arbitrary materials \cite{22,23,24}. However, the obtaining of
exact formulas for the Casimir free energy and force between
bodies with nonplanar surfaces was challenging during the last few
decades. The problem was solved in 2006 when several equivalent
Lifshitz-type formulas were obtained \cite{25,26,27,28} using the
scattering theory and the formalism of functional determinants.
Specifically, in one of the developed formalisms the Casimir
free energy for two separated bodies of arbitrary shape
labeled by the indices (1) and (2) can be presented in the
form \cite{29}
\begin{equation}
{\cal F}^{\rm exact}(a,T)=k_BT\sum_{l=0}^{\infty}
{\vphantom{\sum}}^{\prime}\ln\,{\rm det}
\left[1-{\cal T}^{(1)}{\cal G}_{\xi_l;(1,2)}^{(0)}
{\cal T}^{(2)}{\cal G}_{\xi_l;(2,1)}^{(0)}\right].
\label{eq8}
\end{equation}
\noindent
Here, ${\cal G}_{\xi_l;(1,2)}^{(0)}$ is the operator for the free
space Green's function with the matrix elements
 $\langle \mbox{\boldmath$r$}|{\cal G}_{\xi_l;(1,2)}^{(0)}|
\mbox{\boldmath$r$}^{\prime}\rangle $,
where $\mbox{\boldmath$r$}$ belongs to the body (1) and
$\mbox{\boldmath$r$}^{\prime}$ belongs to the body (2).
${\cal T}^{(1)}$ [${\cal T}^{(2)}$] is the operator of $T$ matrix for
the body (1) [body (2)]. Using such representations, it was shown
that for an ideal metal sphere above an ideal metal plate
at room temperature the exact results for the Casimir force and for
the thermal correction to it coincide with respective results
obtained using the PFA in the zeroth order of the small parameter
$a/R$ \cite{29a}.

Calculation of the Casimir force between real material bodies using
the Lifshitz theory requires the availability of the respective
$\varepsilon(\omega)$ and $\mu(\omega)$. We reserve the discussion
of $\mu(\omega)$ for Sec.~4 and briefly consider here the dielectric
permittivity for dielectrics and metals. It is common knowledge
that all condensed bodies are either dielectrics or metals
depending on the behavior of their dc conductivity $\sigma_0(T)$
when temperature vanishes \cite{30,31}. For dielectrics
$\sigma_0(0)=0$ whereas for metals $\sigma_0(0)\neq 0$.
In this regard semiconductors are also either dielectrics or
metals (specifically, intrinsic semiconductors are dielectrics,
whereas doped semiconductors with dopant concentration above
the critical value are metals).
In the application of the Lifshitz theory to dielectric materials
the small dc conductivity arising at $T\neq 0$ is usually
neglected. In so doing the dielectric permittivities of
dielectric plates are determined by the core electrons \cite{32}
\begin{equation}
\varepsilon_c^{(n)}(\omega)=1+\sum_{j=1}^{K}
\frac{g_j^{(n)}}{{\omega_j^{(n)}}^2-\omega^2-i\gamma_j^{(n)}\omega},
\label{eq9}
\end{equation}
\noindent
where $\omega_j^{(n)}\neq 0$ are the oscillator frequencies,
$g_j^{(n)}$ are the oscillator strengths, $\gamma_j^{(n)}$
are the damping parameters, and $K$ is the number of oscillators.
Note that the parameters of
oscillators in Eq.~(\ref{eq9}) can be temperature-dependent.
Thus, in addition to the explicit dependence on the temperature in
Eqs.~(\ref{eq1}) and (\ref{eq5}) through the multiple $T$ and
through the Matsubara frequencies, there is an implicit
dependence through the dielectric permittivity. For Si such
an implicit dependence was investigated in Ref.~\cite{32a}.

However, if one takes into account the existence of free charge
carriers at $T\neq 0$, the dielectric permittivity
of dielectric plates can be represented as \cite{33}
\begin{equation}
\varepsilon_d^{(n)}(\omega)=\varepsilon_c^{(n)}(\omega)+
i\frac{4\pi\sigma_0^{(n)}(T)}{\omega}.
\label{eq10}
\end{equation}
\noindent
It has been shown \cite{34,35} that the Casimir entropy calculated
using the free energy (\ref{eq1}) combined with the dielectric
permittivity (\ref{eq9}) satisfies an important thermodynamic
constraint
\begin{equation}
S_c(a,T)=-\frac{\partial{\cal F}_c(a,T)}{\partial T}\to 0
\label{eq11}
\end{equation}
\noindent
when $T\to 0$ (the Nernst heat theorem). If, however, the dc
conductivity of dielectric material is taken into account
[i.e., the dielectric permittivity (\ref{eq10}) is used]
we get \cite{34,35}
\begin{equation}
S_d(a,0)=\frac{k_B}{16\pi a^2}\left[\zeta(3)-
{\rm Li}_3(r_0^{(1)}r_0^{(2)})\right]>0,
\label{eq12}
\end{equation}
\noindent
where
\begin{equation}
r_0^{(n)}=\frac{\varepsilon_c^{(n)}(0)-1}{\varepsilon_c^{(n)}(0)+1}<1,
\label{eq12a}
\end{equation}
\noindent
$\zeta(z)$ is the Riemann zeta function, and ${\rm Li}_n(z)$ is
the polylogarithm function. Keeping in mind that real dielectric
materials possess dc conductivity, one arrives to the conclusion
that for two parallel dielectric semispaces the Lifshitz theory
violates the Nernst teorem.

A similar difficulty arises in the application of the Lifshitz theory
to real metals. It is customary to describe the dielectric properties
of metallic plates by means of the Drude model
\begin{equation}
\varepsilon_D^{(n)}(\omega)=1-
\frac{{\omega_p^{(n)}}^2}{\omega[\omega+i\gamma^{(n)}(T)]}\, ,
\label{eq13}
\end{equation}
\noindent
where $\omega_p^{(n)}$ are the plasma frequencies and
$\gamma^{(n)}(T)$ are the relaxation parameters.
This model is well applicable in the wide frequency region from
quasistatic frequencies (the normal skin effect) to infrared
optics. It was shown \cite{6,7a,36,37} that for perfect crystal
lattices [where $\gamma^{(n)}(T)\to 0$ as $T^2$] the Casimir
entropy calculated using the free energy (\ref{eq1}) combined
with the Drude model (\ref{eq13}), violates the Nernst heat
theorem
\begin{equation}
S_D(a,0)=-\frac{k_B\zeta(3)}{16\pi a^2}\left[1-
\frac{2c}{a}\,
\frac{\omega_p^{(1)}+\omega_p^{(2)}}{\omega_p^{(1)}\omega_p^{(2)}}+
\frac{3c^2}{a^2}\,\left(
\frac{\omega_p^{(1)}+\omega_p^{(2)}}{\omega_p^{(1)}\omega_p^{(2)}}
\right)^2-\ldots\right]<0.
\label{eq14a}
\end{equation}
\noindent
Notice that for Drude metals with impurities there is a nonzero
residual relaxation $\gamma_{\rm res}^{(n)}=\gamma^{(n)}(0)$.
In this case the Casimir entropy jumps abruptly to zero at
$T<10^{-4}\,$K starting from a negative value (\ref{eq14a}), so
that the Nernst theorem is formully satisfied \cite{38,39,40}.
This, however, does not solve the problem. The point is that
a perfect crystal lattice is a truly equilibrium system with
a nondegenerate dynamical state of the lowest energy. Thus,
according to quantum statistical physics, the Casimir entropy
calculated for a perfect crystal lattice must be equal to zero
\cite{41}.

It is quite another matter when the nondissipative plasma model
\begin{equation}
\varepsilon_p^{(n)}(\omega)=1-
\frac{{\omega_p^{(n)}}^2}{\omega^2}\, ,
\label{eq14}
\end{equation}
\noindent
is used instead of (\ref{eq13}) to describe metallic plates.
In this case the Nernst heat theorem is satisfied
\cite{6,7a,36,37}, i.e.
\begin{equation}
S_p(a,T)=-\frac{\partial{\cal F}_p(a,T)}{\partial T}\to 0
\label{eq15}
\end{equation}
\noindent
when $T\to 0$. It was proposed \cite{6,7a,42} that the inclusion
of relaxation properties of free charge carriers in the model of
dielectric response violates thermal equilibrium which is the
basic applicability condition of the Lifshitz theory. This is
because the drift electric current leads to a violation of
time-reversal symmetry and requires the introduction of the
unidirectional flux of heat from the Casimir plates to the heat
reservoir to preserve the temperature constant. Consensus on the
problem of why the Lifshitz theory is thermodynamically inconsistent
in the presence of free charge carriers is not yet achieved.

To complete a brief survey of the Lifshitz theory, we mention
recent attempts to modify the TM reflection coefficients by
taking into account the effect of screening \cite{43,44}.
This approach includes both the drift and diffusion currents of
free charge carriers. Eventually  the modified TM reflection
coefficient is obtained through the use of Boltzmann transport
equation and takes the form \cite{44}
\begin{equation}
r_{\rm TM}^{(n)\rm mod}(ix,k_{\bot})=
\frac{\varepsilon^{(n)}(ix)q-k^{(n)}-\frac{k_{\bot}^2}{\eta^{(n)}(ix)}\,
\frac{\varepsilon^{(n)}(ix)-
\varepsilon_c^{(n)}(ix)}{\varepsilon_c^{(n)}(ix)}}{\varepsilon^{(n)}(ix)q
+k^{(n)}+\frac{k_{\bot}^2}{\eta^{(n)}(ix)}\,
\frac{\varepsilon^{(n)}(ix)-
\varepsilon_c^{(n)}(ix)}{\varepsilon_c^{(n)}(ix)}}\, .
\label{eq16}
\end{equation}
\noindent
Here, the dielectric permittivities of the plates are given by
\begin{equation}
\varepsilon^{(n)}(ix)=\varepsilon_c^{(n)}(ix)+
\frac{{\omega_p^{(n)}}^2}{x(x+\gamma^{(n)})},
\label{eq17}
\end{equation}
\noindent
where permittivities of the bound core electrons
$\varepsilon_c^{(n)}$ are defined in Eq.~(\ref{eq9}).
The quantity $\eta^{(n)}(ix)$ is given by
\begin{equation}
\eta^{(n)}(ix)=\left[k_{\bot}^2+{\kappa^{(n)}}^2\,
\frac{\varepsilon_c^{(n)}(0)}{\varepsilon_c^{(n)}(ix)}\,
\frac{\varepsilon^{(n)}(ix)}{\varepsilon^{(n)}(ix)-
\varepsilon_c^{(n)}(ix)}\right]^{1/2} ,
\label{eq18}
\end{equation}
\noindent
where $1/\kappa^{(n)}$ are the screening lengths of plate
materials \cite{45} defined differently for dielectrics
and metals (the Debye-H\"{u}ckel and Tomas-Fermi screening
lengths when the charge carriers are described by the
Maxwell-Boltzmann and Fermi-Dirac statistics, respectively).
The TE reflection coefficient in the proposed approach
remains unchanged.

It was shown \cite{42,46,47,47a}, however,
that the modification of the TM reflection
coefficient in accordance with Eq.~(\ref{eq16}) also leads to the
violation of the Nernst theorem for all dielectric materials
whose charge carrier density does not vanish when $T\to 0$ and
$\sigma_0(T)$ vanishes due to the vanishing mobility
(for instance, for glasses with ionic conductivity \cite{47b}
or doped semiconductors with dopant concentration below the
critical value \cite{47c}). For metals with perfect crystal
lattices the Lifshitz theory with the modified TM reflection
coefficient violates the Nernst theorem in the same way as
when the standard Drude model approach is used \cite{47,48,48a}.
These conclusions were disputed in Refs.~\cite{49,50}, but the
argumentation used was shown to be inadequate in
Refs.~\cite{7,47}.
Therefore, the problem of consistency of the Lifshitz theory
with thermodynamics for real materials remains unsolved.

\section{Comparison with three landmark experiments}

There are about 25 experiments on measuring the Casimir force
in the modern stage of the work on the subject (see the review
\cite{6} for more details). In almost all experiments at least
an attempt of comparing the measurement data with theory was
undertaken. In this section, we consider only three experiments
each of which, when carefully compared with the Lifshitz theory,
leads to unexpected conclusions.

\subsection{Measuring the Casimir pressure using a micromachined
oscillator}

\begin{figure}[t]
\begin{center}
\vspace*{-0.5cm}
\hspace*{-2.cm}
\includegraphics[width=15cm]{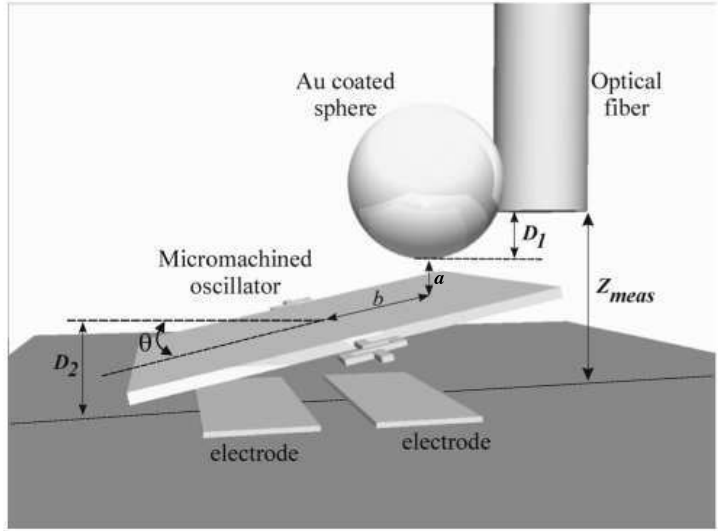}
\vspace*{-19.7cm}
\end{center}\caption{Schematic diagram of the setup for measuring the
gradient of the Casimir force between a sphere and a plate
(which is also an indirect measurement of the Casimir pressure between
two parallel plates) using a micromachined oscillator.}
\end{figure}

The first of this landmark experiments was performed by
R.~S.~Decca and his co-workers. It was repeated three times,
in 2003 \cite{8}, 2005 \cite{9}, and in 2007 \cite{10,11} with
major improvements in sample preparation and in the experimental
precision in each repetition. The discussion below is related to
the most precise experiment of 2007 \cite{10,11}, where the
gradient of the Casimir force between an Au-coated sphere of
$R=151.3\,\mu$m radius and a plate of micromachined oscillator
was measured at $T=300\,$K
in the dynamic regime (see Fig.~1 for the schematic
diagram of a setup). In fact it was the shift of the resonant
frequency of a micromachined oscillator that was measured.
This shift is proportional to the gradient of the Casimir force
between a sphere and a plate. Considering the negative derivative
with respect to $a$ of both sides of Eq.~(\ref{eq7}), one obtains
\begin{equation}
P(a,T)\equiv
-\frac{\partial{\cal F}(a,T)}{\partial a}=
-\frac{1}{2\pi R}\,\frac{\partial{F}(a,T)}{\partial a}.
\label{eq19}
\end{equation}
\noindent
Thus, the use of the PFA allows interpretation of the experiment
by Decca et al. as an indirect measurement of the Casimir
pressure between two parallel plates. In this experiment all
errors were determined at a 95\% confidence level, and the random
errors were made much less than the systematic errors, so that
only the latter determined the total experimental error.
As a result, the total relative experimental error of the Casimir
pressure varied from 0.19\% at $a=162\,$nm to 0.9\% at
$a=400\,$nm and to 9.0\% at the largest separation $a=746\,$nm.
This includes the error of the PFA which was used to recalculate
the experimental data for the gradient of the Casimir force into
the Casimir pressure. In the most conservative way the error of
the PFA was taken to be equal to $a/R$ (specially performed
experiment \cite{51} and theoretical computations \cite{52,52a,53}
for a sphere
and a plate made of real metals described by the plasma and
Drude models show that this error is in fact smaller than
$a/R$).

Computations of the Casimir pressure between Au plates was
performed by Eq.~(\ref{eq1}) with $\mu^{(n)}=1$
using the tabulated optical data \cite{33} for the
complex index of refraction
$n(\omega)={\rm Re}\,n(\omega)+i{\rm Re}\,n(\omega)$.
For this purpose the imaginary part of the dielectric permittivity
of Au was found as
${\rm Im}\,\varepsilon(\omega)=2{\rm Re}\,n(\omega)
{\rm Im}\,n(\omega)$ at frequencies where the optical data are
available and extrapolated to lower frequencies by means of the
Drude model (\ref{eq13}) with the parameters
$\omega_p=9.0\,$eV, $\gamma=0.035\,$eV \cite{54}.
Then $\varepsilon(i\xi)$ for Au was obtained using the
Kramers-Kronig relation
\begin{equation}
\varepsilon(i\xi)=1+\frac{2}{\pi}\int_{0}^{\infty}
\frac{\omega\,{\rm Im}\,\varepsilon(\omega)}{\xi^2+\omega^2}\,
d\omega.
\label{eq20}
\end{equation}
\noindent
This calculational approach corresponds to the use of the Drude
model, as described in Sec.~2, but takes into account the role
of core electrons. Keeping in mind that problems discussed in
Sec.~2 originate from the zero-frequency term of the Lifshitz
formula, one might expect that related difficulties should
appear here as well. For the comparisom with the experimental data,
the surface roughness on both the sphere and the plate was studied
with the help of an atomic force microscope and taken into account
in a nonmultiplicative way by means of geometrical averaging
\cite{6,7a}.

Another approach to calculate the Casimir pressure between real
metals with account of core electrons uses the generalized
plasma-like model for the permittivity
\begin{equation}
\varepsilon(\omega)=\varepsilon_c(\omega)-
\frac{\omega_p^2}{\omega^2},
\label{eq21}
\end{equation}
\noindent
where $\varepsilon_c(\omega)$ is defined in Eq.~(\ref{eq9}) and
the parameters of oscillators for Au with $K=6$ were determined
in Ref.~\cite{11}. The surface roughness is taken into account
as described above.

\begin{figure}[b]
\begin{center}
\vspace*{-7.5cm}
\hspace*{-3.cm}
\includegraphics[width=19cm]{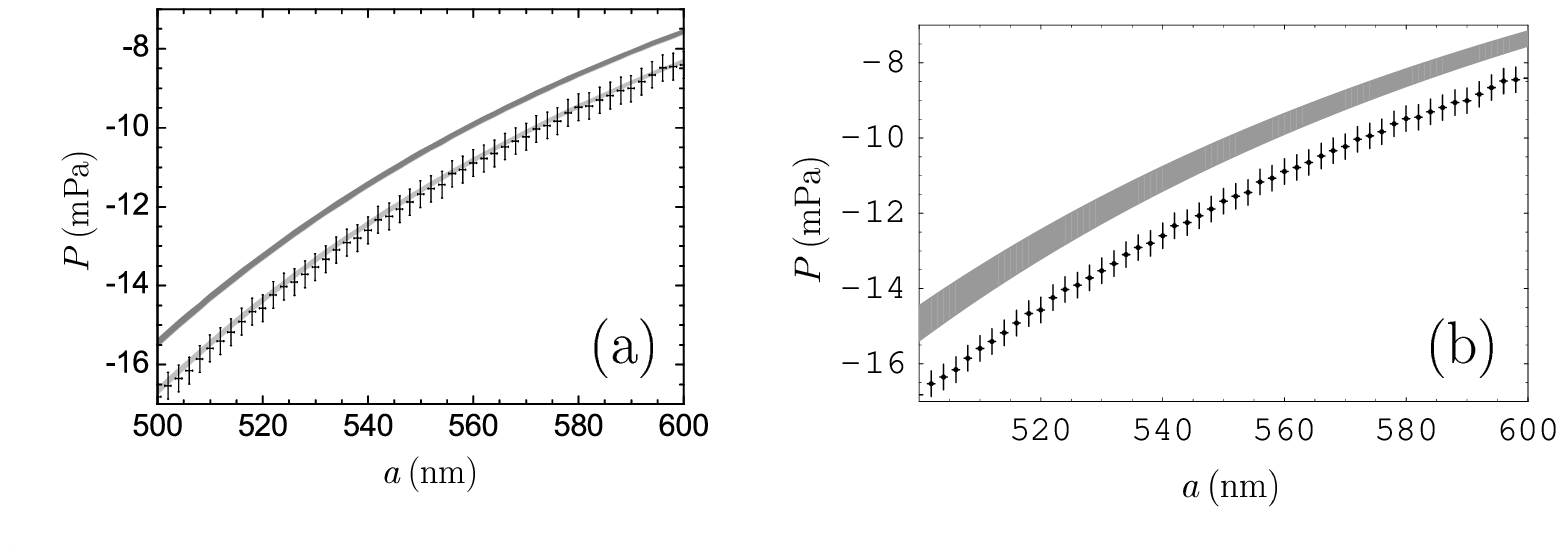}
\vspace*{-15.3cm}
\end{center}\caption{The measured mean Casimir pressure together with
the absolute errors
in the separation and pressure versus separation
is shown as crosses. Errors are calculated at a 95\% confidence
level.
(a) The theoretical Casimir pressure computed using
the generalized plasma-like model and the optical data extrapolated by
the Drude model is shown by the light-gray and dark-gray bands,
respectively. (b) The theoretical Casimir pressure
computed using different sets
of optical data available in the literature versus separation is shown
as the dark-gray band.}
\end{figure}
Now we present a typical comparison between the experimental data
and theory. In Fig.~2(a,b) the mean Casimir pressure together with
the absolute experimental errors in the separation and pressure
measured at different separations is shown as crosses.
In Fig.~2(a) the theoretical Casimir pressure computed using the
generalized plasma-like model and the optical data of
Ref.~\cite{33} extrapolated by the Drude model is shown by the
light-gray and dark-gray bands, respectively. The width of the
bands indicates the total theoretical error. As can be seen in
Fig.~2(a), the plasma-like model is consistent with the data,
whereas a theory using the extrapolation to low frequencies
by means of the Drude model is excluded by the data at a 95\%
confidence level. As was noted in the literature \cite{55,56},
Au films prepared using different technology may possess
different sets of optical data leading to the values of
$\omega_p$ varying from 6.85 to 9.0\,eV. In Fig.~2(b) the
theoretical Casimir pressure as a function of separation
computed with all sets of optical data available is shown as
the dark-gray band. It is seen that the use of any alternative
value of $\omega_p$ makes the disagreement between the Drude
model approach and the data even more evident.
One more recent experiment \cite{57} incorporated the
determination of the Casimir pressure by means of a
micromachined oscillator with simultaneous measurement of
the optical data of Au films. It was shown that while the
Casimir pressures coincide with those measured previously
in Refs.~\cite{10,11}, the optical data closely follow those
tabulated in Ref.~\cite{33}.

\begin{figure}[t]
\begin{center}
\vspace*{-8.cm}
\hspace*{-2.cm}
\includegraphics[width=19cm]{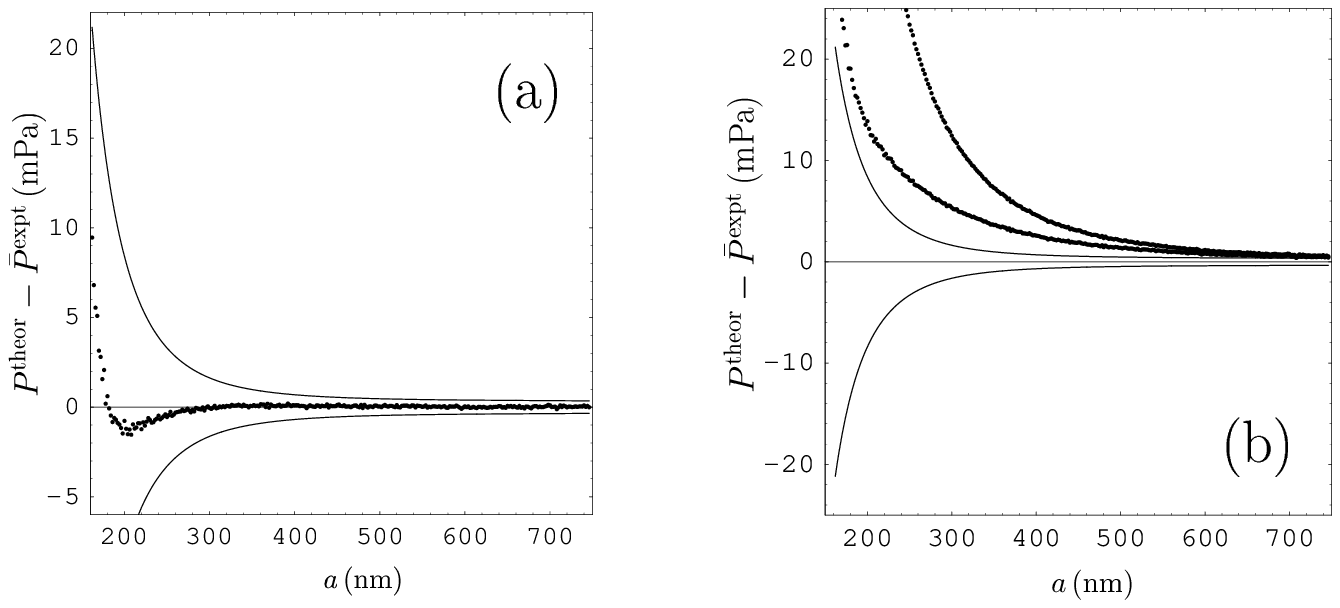}
\vspace*{-14.5cm}
\end{center}\caption{Differences between the theoretical Casimir
pressures, computed by means of (a) the
generalized plasma-like model and (b)
the Drude model approach, and mean experimental Casimir pressures at
different separations are shown as
dots. The lower and upper sets of dots for the Drude model approach
correspond to the use of maximum and minimum values of the plasma
frequency  available in the literature.
The solid lines indicate the borders of the confidence intervals
found at a 95\% confidence level.}
\end{figure}
Another method for comparing theory with experiment considers the
differences between theoretical and mean experimental Casimir
pressures, $P^{\rm theor}(a_i)-\bar{P}^{\rm expt}(a_i)$, and
the 95\% confidence interval
$[-\Xi_P(a),\Xi_P(a)]$ to which 95\% of these differences
must belong if the theory is consistent with the data.
In Fig.~3(a), the differences of the theoretically computed
pressures using the
generalized plasma-like model and mean experimental Casimir
pressures are shown as dots. The solid lines are formed by the
ends of the confidence intervals. As can be seen in Fig.~3(a),
the Lifshitz theory using the generalized plasma-like model
is consistent with the data.
A similar comparison is made
with the theoretical Casimir pressures computed using the
tabulated optical data extrapolated by the Drude model.
In Fig.~3(b) the differences
$P^{\rm theor}(a_i)-\bar{P}^{\rm expt}(a_i)$
are indicated as two sets of dots computed using
$\omega_p=6.82\,$eV (the upper set) and
$\omega_p=9.0\,$eV (the lower set).
The pressure differences for all other values of the plasma
frequency of Au considered in Ref.~\cite{56} are
between the two sets of dots in Fig.~3(b). Thus, theory
using the Drude model is experimentally excluded over the
entire measurement range from 162 to 746\,nm at a 95\%
confidence level. Note that within a more narrow range of
separations from 210 to 620\,nm the theoretical approach
using the Drude model is excluded by the data at a 99.9\%
confidence level \cite{6,7a,11}. The same data exclude
\cite{57a} the Lifshitz theory at zero temperature combined with
the Drude model. This exclusion is made at a 70\% confidence
level with the optical data of Ref.~\cite{33} and at 95\% confidence
level with the alternative optical data.

The experimental data of Ref.~\cite{10,11} have been also compared
with theoretical predictions obtained using the modified TM
reflection coefficient (\ref{eq16}). In so doing it was supposed
that electrons in Au are described by Fermi-Dirac statistics.
Almost the same computational results for the Casimir pressure
were obtained \cite{47,48,48a}, as by using the Drude model.
Thus, the Lifshitz theory with the modified TM reflection
coefficient is excluded by the experiment of Refs.~\cite{10,11}.

\subsection{Optical modulation of the Casimir force}

The second landmark experiment was performed by U.~Mohideen
and his co-workers \cite{12,13}. In this experiment, an atomic
force microscope was employed to measure the change in the
Casimir force between an Au-coated sphere ($R=98.9\,\mu$m) and
a Si membrane at $T=300\,$K in the presence and in the absence
of incident laser light on a membrane (see schematic of the
experimental setup in Fig.~4).
\begin{figure}[t]
\begin{center}
\vspace*{-2.cm}
\hspace*{-1.5cm}
\includegraphics[width=18cm]{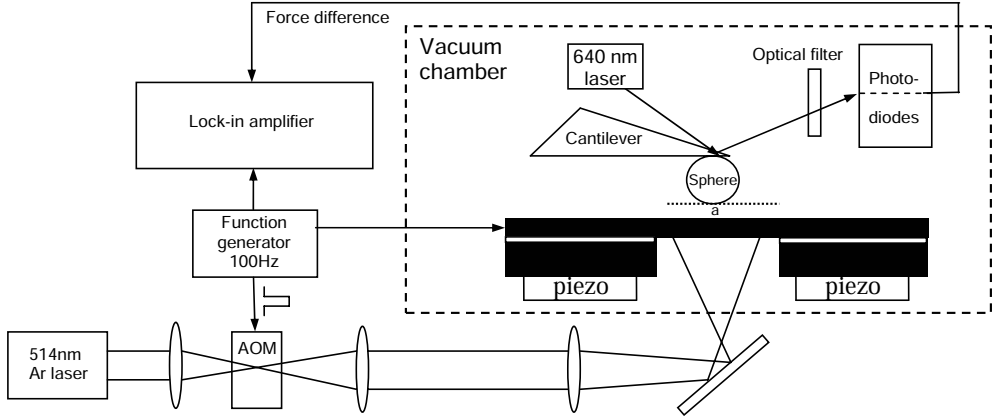}
\vspace*{-15.5cm}
\end{center}\caption{Schematic of the experimental setup on measuring
the difference Casimir force between Au sphere and Si plate illuminated
with laser pulses.}
\end{figure}
The experiment was repeated three times with three different
absorbed laser powers. The values of the maximum and minimum absorbed
powers were equal to $P_a=9.3\,$mW and $P_b=4.7\,$mW.
In the absence of laser light, the charge carrier density of
high-resistivity Si was
$n_h=5\times 10^{14}\,\mbox{cm}^{-3}$.
In the presence of light, the charge carrier densities related to
the maximum and minimum absorbed powers took the values
$n_a=(2.1\pm 0.4)\times 10^{19}\,\mbox{cm}^{-3}$ and
$n_b=(1.4\pm 0.3)\times 10^{19}\,\mbox{cm}^{-3}$ (the errors are
indicated at a 95\% confidence level \cite{13}).

The difference of the theoretical Casimir forces in the presence
and in the absence of laser light,
\begin{equation}
F_{\rm diff}^{\rm theor}(a,T)=F_{l}^{\rm theor}(a,T)-
F^{\rm theor}(a,T),
\label{eq22}
\end{equation}
\noindent
was calculated using Eq.~(\ref{eq1}) with $\mu^{(n)}=1$ and
Eq.~(\ref{eq7}), taking into account the surface roughness by
means of geometrical averaging. The dielectric permittivity of
the Au-coated sphere
$\varepsilon^{(1)}(i\xi)\equiv\varepsilon_{\rm Au}(i\xi)$ is
described by Eq.~(\ref{eq21}). The dielectric permittivity of the Si
plate $\varepsilon^{(2)}(i\xi)\equiv\varepsilon_{\rm Si}(i\xi)$
depends on the concentration of charge carriers. In the absence
of charge carriers (infinitely high resistivity) it was computed
by using the tabulated optical data \cite{33} and the
Kramers-Kronig relation. The obtained results,
$\varepsilon_{{\rm Si},0}(i\xi)$, are shown by the dashed line
in Fig.~5(a). The dielectric permittivity of high-resistivity
plate used in the experiment in the absence of laser light,
$\varepsilon_{{\rm Si},h}(i\xi)$,  can be presented as the sum of
$\varepsilon_{{\rm Si},0}(i\xi)$ and the second term on the
right-hand side of Eq.~(\ref{eq10}) taking the dc conductivity
into account. In Fig.~5(a) the quantity
$\varepsilon_{{\rm Si},h}(i\xi)$ is shown by the dotted line.
Similarly, the dielectric permittivity of low-resistivity Si with
different absorbed powers, $\varepsilon_{{\rm Si},a,b}(i\xi)$,
are shown by the solid lines a and b taking into account both
electrons and holes \cite{58,59,59a}.
\begin{figure}[t]
\begin{center}
\vspace*{-1.5cm}
\hspace*{-5.cm}
\includegraphics[width=24cm]{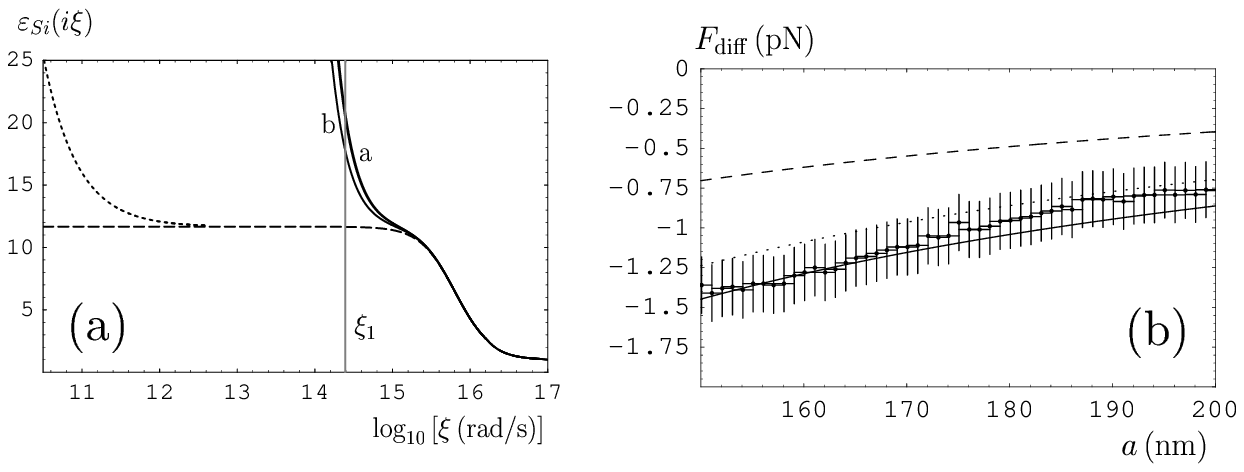}
\vspace*{-28.7cm}
\end{center}\caption{
(a) The dielectric permittivity of the Si membrane
along the imaginary frequency axis for Si with different concentration
of charge carriers (see text for further discussion). The vertical
gray line indicates the first Matsubara frequency at $T=300\,$K.
(b) The experimental differences in the Casimir
force with their experimental errors are shown as crosses (the absorbed
power is equal to 4.7\,mW). The solid and dotted lines
 represent the theoretical differences
computed at $T=300\,$K using the model with a finite static
permittivity of high-resistivity Si, but different models for Si in
the presence of light (see text for further discussion).
The dashed line represents the theoretical differences
computed  including the dc
conductivity of high-resistivity Si.}
\end{figure}

Now we are in a position to compare the experimental data with the
predictions of the Lifshitz theory. In Fig.~5(b) the mean
experimental differences, $\bar{F}_{\rm diff}^{\rm expt}$, of the
Casimir forces in the presence and in the absence of laser light
for the absorbed power $P_b$ are shown as crosses. The arms of
the crosses represent total experimental  errors in the
measurement of the difference force and separation distance,
determined at a 95\% confidence level. The theoretical differences
of the Casimir force (\ref{eq22}) were first computed using the
dielectric permittivity $\varepsilon_{{\rm Si},h}(i\xi)$ in the
absence of light which takes the dc conductivity of a Si plate
into account. In the presence of light the permittivity
$\varepsilon_{{\rm Si},b}(i\xi)$ has been used. The computational
results are presented by the dashed line in Fig.~5(b). As can be
seen in the figure, the prediction of the Lifshitz theory taking
into account the dc conductivity of high-resistivity Si is
experimentally excluded. Then the theoretical differences
(\ref{eq22}) were computed using the dielectric permittivity
$\varepsilon_{{\rm Si},0}(i\xi)$ in the dark phase which
disregards the dc conductivity of Si. In the presence of light
the permittivity $\varepsilon_{{\rm Si},b}(i\xi)$ was again
used. In this case the computational results are shown by the
solid line in Fig.~5(b) which is consistent with the data.

Note that the dielectric permittivity
$\varepsilon_{{\rm Si},b}(i\xi)$ used in the presence of light
describes charge carriers in the low-resistivity Si by means
of the Drude model. If to take into account that $n_b$ is
larger than the critical value of charge carrier density
indicating the dielectric to metal phase transition, it would
be reasonable to use the description by means of the generalized
plasma-like model in the presence of light (see Sec.~2).
The computational results obtained within this approach \cite{60}
are presented by the dotted line in Fig.~5(b). It is seen that
the dotted line is consistent with the data, as well as the solid
line. Thus, this experiment is not of sufficient precision to
discriminate between the description of charge carriers in
metallic phase by means of the Drude and plasma models (this was
done by the experiment of Decca et al. considered in Sec.~3.1).
The main new and unexpected result of the experiment by
Mohideen et al. is that the inclusion of the dc conductivity
in a dielectric phase (high-resistivity Si) results in the
contradiction of the Lifshitz theory with the data. Because
of this, phenomenologically one may conclude \cite{35} that
in applications of the Lifshitz theory to dielectric materials
the dc conductivity should be disregarded.
Note that the contribution of dc conductivity to the dielectric
permittivity of dielectrics was also disregarded when studying
nonequilibrium thermal fluctuations of the electric field
in the presence of a dc current \cite{69a}.

\begin{figure}[t]
\begin{center}
\vspace*{-7.5cm}
\hspace*{-2.3cm}
\includegraphics[width=19cm]{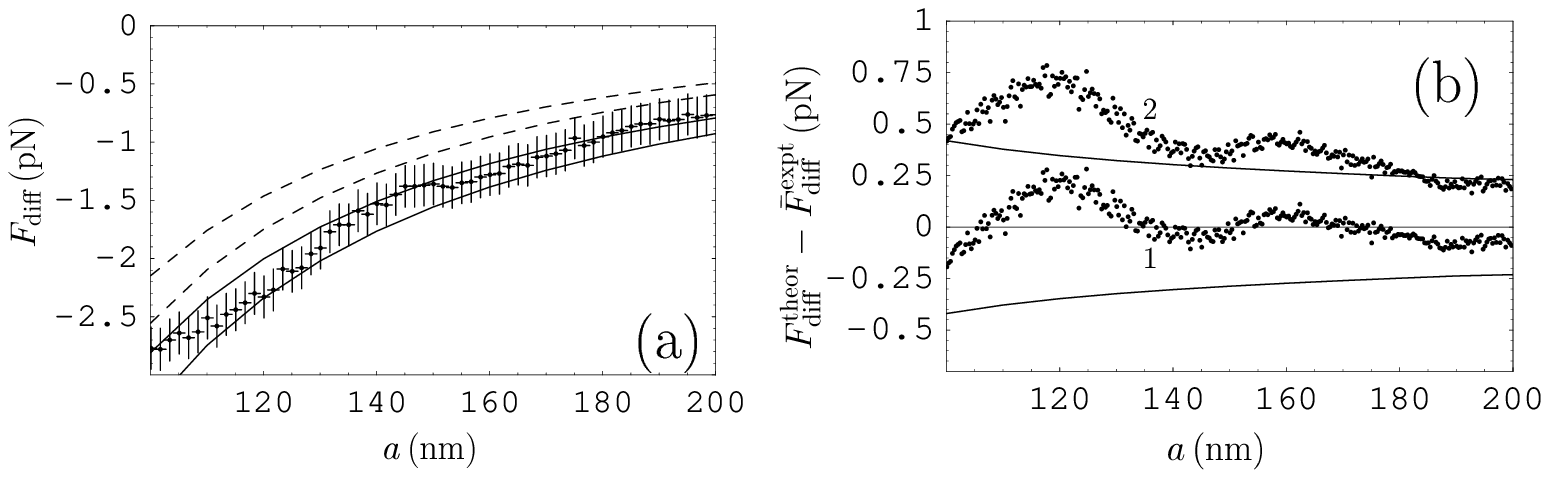}
\vspace*{-15.3cm}
\end{center}\caption{
(a) The measured force differences are shown as crosses
versus separation $a$.
The two theoretical bands lie between the solid and dashed lines
(see text for details).
(b) Theoretical minus mean experimental differences of the
Casimir force versus separation are shown as dots.
For the dots labeled 1, the theoretical results  are computed with
a finite static dielectric permittivity of high resistivity Si
in the absence of laser pulse  and using  the plasma model
to describe  the charge carriers when the laser pulse is on.
For dots labeled 2, the theoretical results are computed
in the framework of the modified Lifshitz theory \cite{43,44}.
The solid lines
indicate 70\% confidence intervals including all experimental and
theoretical errors.}
\end{figure}
The experimental data of the optical modulation experiment was
compared with the modified Lifshitz theory using the TM
reflection coefficient (\ref{eq16}) proposed in
Refs.~\cite{43,44}. It was found that predictions of the
modified Lifshitz theory for the Si plate in a dielectric phase
cannot be excluded at a 95\% confidence level. Because of this,
the comparison has been made at a 70\% confidence level
\cite{42}. In Fig.~6(a) the experimental data for the measured
difference Casimir force in the presence and in the absence of
light are shown as crosses. When compared to Fig.~5(b), the
arms of the crosses are made relatively shorter, as required by
a lower confidence level. The prediction of the modified
Lifshitz theory for the absorbed power $P_b$ is shown as a band
bounded by the two dashed lines.
The width of this band is found from the theoretical error
determined mostly by the error in $n_b$ (at a 70\% confidence
level this error is equal to $0.22\times 10^{19}\,\mbox{cm}^{-3}$
to compare with $0.3\times 10^{19}\,\mbox{cm}^{-3}$ at a
95\% confidence level). The prediction of the standard Lifshitz
theory with dc conductivity of Si in the absence of light
disregarded is shown by the band bounded by two solid lines.
The width of this band also corresponds to an error in $n_b$
determined at a 70\% confidence level. As can be seen in Fig.~6(a),
the predictions of the modified Lifshitz theory \cite{43,44} are
excluded by the data at a 70\% confidence level, whereas the
standard Lifshitz theory with disregarded dc conductivity of
dielectric Si is consistent with the data.

The same conclusion can be obtained when using another approach to
the comparison between experiment and theory similar to that in
Fig.~3. In Fig.~6(b) the differences between theoretical and mean
experimental difference Casimir forces,
$F_{\rm diff}^{\rm theor}-\bar{F}_{\rm diff}^{\rm expt}$,
at different separations are plotted as the two sets of dots.
The set labeled 1 is computed using the
standard Lifshitz theory with the dc conductivity of
dielectric Si disregarded. The set labeled 2 is computed using
the modified Lifshitz theory. The two solid lines indicate the
borders of a 70\% confidence intervals determined with account
of all experimental and theoretical errors. It is seen that the
standard Lifshitz theory with the dc conductivity  disregarded
is consistent with the data, whereas the modified Lifshitz
theory \cite{43,44} is excluded by the measurement results of
the optical modulation experiment at a 70\% confidence level.

\subsection{Measurements of the thermal Casimir-Polder force}

The third landmark experiment is on measuring the Casimir-Polder
force between ground state ${}^{87}$Rb atoms and a fused
silica plate. It was performed by Cornell and his co-workers
\cite{14}. Atoms of ${}^{87}$Rb belonged to the Bose-Einstein
condensate separated by a distance of a few micrometers from
a wall. The experiment was of dynamic-type. The dipole
oscillations in the direction perpendicular to the plate
with the frequency $\omega_0$ were excited in a condensate,
and the relative shift of this frequency
\begin{equation}
\gamma_z=\frac{|\omega_0-\omega_z|}{\omega_0}
\label{eq23}
\end{equation}
\noindent
due to the Casimir-Polder interaction was measured precisely.
This shift can be recalculated \cite{62} into the
Casimir-Polder force (\ref{eq5}) with $\mu^{(n)}=1$,
$\beta(i\xi)=0$ leading to its indirect measurement.

This experiment was repeated three times, once in
thermal equilibrium, when the temperature of the plate was equal
to the environment temperature, $T_p=T_e=310\,$K, and
two times with $T_p>T_e$ (in the latter case there are additions
to the Casimir-Polder force (\ref{eq5}) calculated in
Ref.~\cite{63}). The relative frequency shift
(\ref{eq23}) was measured
at several atom-plate separations in the region from 7 to
$11\,\mu$m. Keeping in mind that the comparison of the frequency
shift with the Casimir-Polder force includes an averaging over
the condensate cloud, the role of atom-plate separation is in
fact played by the distance between the plate and the center of
mass of the condensate.

\begin{figure}[b]
\begin{center}
\vspace*{-12.5cm}
\hspace*{-3.cm}
\includegraphics[width=22cm]{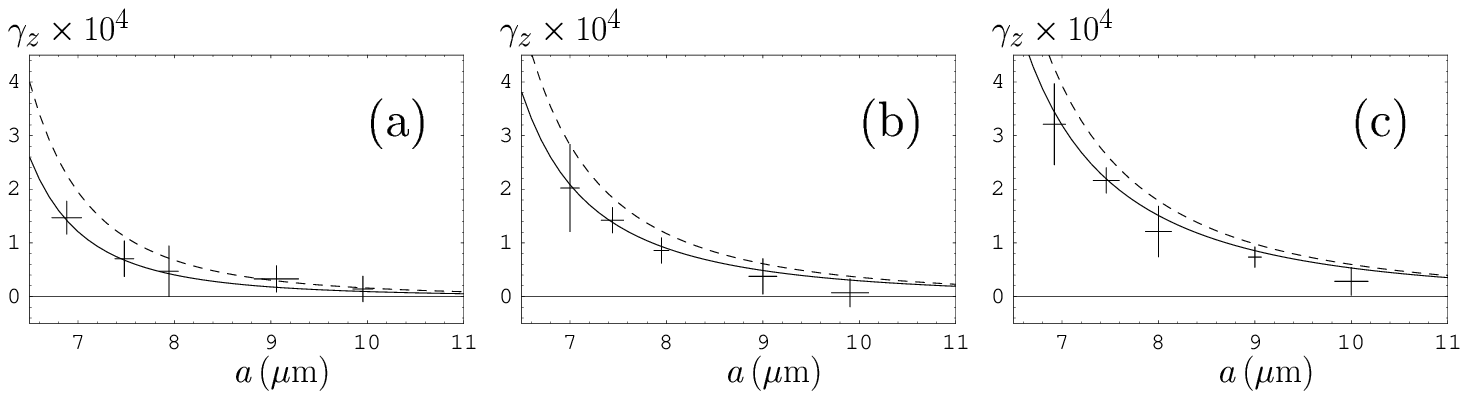}
\vspace*{-15.1cm}
\end{center}\caption{The fractional change in the trap frequency
versus separation (a) in thermal
equilibrium with $T_p=T_e=310\,$K and out  of  thermal equilibrium
(b) with $T_p=479\,$K and $T_e=310\,$K
and (c) with $T_p=605\,$K and $T_e=310\,$K.
Computations are done by neglecting (solid lines)
and including (dashed lines) the dc conductivity of the dielectric
plate. The experimental data are shown as crosses.
}
\end{figure}
In Fig.~7(a) the experimental data for the relative frequency
shift $\gamma_z$ in thermal equilibrium at separations below
$10\,\mu$m are shown as crosses \cite{14}. The arms of the
crosses indicate total experimental errors in the
measurement of $\gamma_z$ and separation distances at a
66\% confidence level found at each experimental point
separately using the analysis of Ref.~\cite{64}.
To compare the experimental data with theory, in Ref.~\cite{14}
the relative frequency shift (\ref{eq23}) was computed using the
Lifshitz formula (\ref{eq5}) for the Casimir-Polder force and an
averaging over the condensate cloud. In so doing the static
value for the polarizability of Rb atom
$\alpha(0)=4.73\times 10^{-23}\,\mbox{cm}^{3}$ can be used.
The dielectric permittivity of fused silica
$\varepsilon(i\xi_l)$, as a function of $\xi_l$, can be
calculated \cite{65} using the tabulated optical data \cite{33}
and the Kramers-Kronig relation (\ref{eq16}).
In the original Ref.~\cite{14} it was assumed that the fused
silica is a true dielectric of infinitely high resistivity.
Thus, the dc conductivity of the plate was disregarded.
Under this assumption the obtained theoretical results \cite{14}
are shown by the solid line in Fig.~7(a). It is seen that the
Lifshitz theory of the Casimir-Polder interaction with the dc
conductivity of fused silica disregarded is consistent with the
measurement data.

It should be noted, however, that although fused silica is a good
insulator, it possesses a nonzero conductivity at $T_p=310\,$K.
This conductivity is ionic in nature and is determined by the
concentration of impurities (alkali ions). In Ref.~\cite{15}
the computations of Ref.~\cite{14} were repeated both
disregarding and including the dc conductivity of fused silica.
When the dc conductivity was disregarded, the results of
Ref.~\cite{14} given by the solid line were reproduced.
When, however, the dc conductivity was included, quite
different results were obtained shown by the dashed line in
Fig.~7(a). It is seen that with inclusion of the dc
conductivity the Lifshitz theory is inconsistent with at least
two experimental points at shortest separations.

The second and third repetitions of the experiment of
Ref.~\cite{14} were performed with the same temperature of
environment, $T_e=310\,$K, but with increased temperatures
of the plate, $T_p=479\,$K and 605\,K, respectively.
The obtained experimental data are shown as crosses in
Fig.~7(b) for $T_p=479\,$K and Fig.~7(c) for $T_p=605\,$K.
The computational results of Ref.~\cite{14} disregarding the
dc conductivity of fused silica are shown as the solid lines
in Fig.~7(b,c). They were reproduced in Ref.~\cite{15}
under an assumption that the dc conductivity of fused silica is
omitted. As is seen in Fig.~7(b,c), the solid lines are in a
good agreement with the experimental data. In Ref.~\cite{15}
computations of the relative frequency shift $\gamma_z$ for
the second and third repetitions of the experiment with
increased temperature of the plate were performed taking the
dc conductivity of fused silica into account.
 The computational results shown
by the dashed lines in Fig.~7(b,c) differ dramatically
from the results obtained with the dc conductivity disregarded.
As can be seen in Fig.~7(b), the three experimental points for
$T_p=479\,$K exclude the dashed line and the other two only
touch it. As to the dashed line in Fig.~7(c), it is seen that all
data points at $T_p=605\,$K exclude the theoretical prediction
computed taking the dc conductivity of fused silica into
account. Thus, measurements of the thermal Casimir force between
${}^{57}$Rb atoms and fused silica plate suggest that in the
application of the Lifshitz theory to dielectric materials the
dc conductivity of these materials should be omitted.

In fact, the results of all three landmark experiments performed
by three different experimental groups are in agreement.
According to these results, the relaxation properties of free
charge carriers are irrelevant to the phenomenon of dispersion
forces and should be disregarded. For metals characterized by a
high density of charge carriers, they should be accounted for
by means of the plasma model. For dielectrics charge carriers
should be ignored. If this phenomenological prescription
\cite{6,7a,35,60} is used, the Lifshitz theory agrees well with
all available experimental data.

\section{Possibilities to control the magnitude and sign of the
Casimir force}

Applications of the Casimir force to nanotechnology require
some operational means to modify the force magnitude
at a fixed separation distance and even change the force
sign with varying separation.
This would permit to realize the actuation of devices
with the Casimir
force at a nanoscale and, when necessary, avoid such undesired
effects as pull-in and stiction mentioned in Sec.~1.
Below we briefly discuss different possibilities to control
the Casimir force proposed in the literature some of which
were already tested experimentally. Special attention is paid to
magnetic materials which could significantly widen the role
of Casimir forces in applied science.
Here, we do not discuss the Casimir force between
metamaterials because the application of the Lifshitz formula to
such materials is questionable, and corresponding experimental
data are not yet available.
In spite of repeated speculations on the possibility of Casimir
repulsion, it was shown \cite{meta1} that the Casimir force
between two plates made of metamaterials and separated by a
vacuum gap is always attractive.

\subsection{Different ways of modifying dielectric properties}

To modify the magnitude of the Casimir force, the dielectric
properties of interacting bodies should be changed over a wide
frequency region. The most apparent possibility of decreasing
the magnitude of the Casimir force, as compared with the
case of two metallic test bodies, is to make one of them
of a semiconductor or dielectric (or at least to cover them
with sufficiently thick dielectric coating). As an example,
in the experiment of Refs.~\cite{66,67} the Casimir force
acting between a Au coated sphere and Si plate at $a=100\,$nm
in vacuum was of about 30\% less than between the same Au
sphere and a Au plate. In the experiment of Ref.~\cite{68} the
Casimir forces acting between a Au sphere and two Si plates
with significantly different dopant concentrations were
sequentially measured. The density of charge carriers in the
first plate was
$n_a\approx 1.2\times 10^{16}\,\mbox{cm}^{-3}$
and in the second plate
$n_b\approx 3.2\times 10^{20}\,\mbox{cm}^{-3}$.
It was found that by choosing plates with such different
concentrations of charge carriers the Casimir force acting
between a Au sphere and a Si plate can vary by 5\% and 8\%
at separations $a=60$ and 100\,nm, respectively.
In Ref.~\cite{69} the measured values of the Casimir force
between Au sphere and Au plate were compared with respective
measurements resulted from a Au sphere and a glass plate
covered with a 190\,nm layer of indium tin oxide which is
a conductor at low frequencies (with respect to the Casimir
force this material was first discussed in Ref.~\cite{36}).
Within the separation region from 50 to 150\,nm the Casimir
force between a sphere and a plate covered with indium tin oxide
was shown to be 40\%--50\% less than between the same
sphere and a Au plate.

The use of test bodies made of dielectric material or coated with
dielectric layers allows to decrease the magnitude of the Casimir
force but does not allow to change this magnitude
periodically. The latter was realized in the
optical modulation experiment of Refs.~\cite{12,13} considered
in Sec.~3.2. In this experiment, the differences of the
Casimir forces in the presence and in the absence of laser light
on a Si plate are equal to 3\% and 6.5\% of the Casimir force
in the absence of light at separations $a=100$ and 150\,nm,
respectively. It is important that here the magnitude of the
Casimir force varies periodically as long as the laser operated
in the pulse mode is switched on.

Optical modulation of the Casimir force in vacuum makes it
possible to periodically change the magnitude of the Casimir
force but does not allow to replace attraction with repulsion.
Considerable opportunities for nanotechnology would be opened
up if pulsating Casimir plates were possible, moving back
and forth under the action of the Casimir force without use of
micromechanical springs. This can be achieved only by using
both attractive and repulsive Casimir forces. In this respect,
it should be noted that there is the case where the Casimir
repulsion is well understood, i.e., the repulsion between the
two plates described by the dielectric permittivities
$\varepsilon^{(1,2)}(i\xi)$ separated with a liquid layer
having the dielectric permittivity $\varepsilon^{(0)}(i\xi)$
such that \cite{5,6,7a}
\begin{equation}
\varepsilon^{(1)}(i\xi)<\varepsilon^{(0)}(i\xi)<
\varepsilon^{(2)}(i\xi).
\label{eq24}
\end{equation}
\noindent
The Casimir repulsion in three-layer systems was observed in
Refs.~\cite{70,71,72,73} (see also Refs.~\cite{74,74a}
concerning additional difficulties arising for measurements
of the Casimir force in liquids).

Recently it was shown \cite{75} that the illumination of one
(Si) plate in three-layer systems Au\,--\,ethanol\,--\,Si,
Si\,--\,ethanol\,--\,Si, and
$\alpha$-Al${}_2$O${}_3$\,--\,ethanol\,--\,Si with laser pulses
can change the Casimir attraction to the Casimir repulsion
and vice versa. This allows to combine the advantages
of the optical modulation with the Casimir repulsion in
three-layer systems. As a result, one can obtain attraction
and repulsion interchangeably at regular intervals.
Calculations show \cite{75} that in the system
Au\,--\,ethanol\,--\,Si the Casimir force is repulsive at
$a>160\,$nm. The illumination of the Si plate with laser light
changes this repulsion for attraction. In the system
Si\,--\,ethanol\,--\,Si, the Casimir force between the Si plates
is attractive in the absence of laser light. However, when
one of the two Si plates is illuminated, the attraction is
replaced with repulsion at separations $a>175\,$nm \cite{75}.
In both these systems the magnitude of
the repulsive pressure is several
times less than the magnitude of the
attractive pressure at the same separation. However, in the
three-layer system $\alpha$-Al${}_2$O${}_3$\,--\,ethanol\,--\,Si,
where the Si plate is illuminated with laser pulses, the
light-induced Casimir repulsion is of the same order of
magnitude as the Casimir attraction. Specifically, in
this system in the presence of light the Casimir repulsion
arises at $a>70\,$nm. Thus, at $a=110\,$nm the attractive
Casimir pressure in the absence of light is equal to
about --75\,mPa, whereas the repulsive pressure in the
presence of light is equal to 50\,mPa.

The dielectric properties undergo changes during different
phase transitions. It was shown \cite{76} that the change
in the Casimir free energy associated with the phase
transition of a metal to the superconducting state is
very small. Nevertheless, according to Refs.~\cite{77,78},
the magnitude of this change is comparable to the condensation
energy of a superconducting film and causes a measurable
increase in the value of the critical magnetic field.
It was also proposed \cite{79} to measure the change of the
Casimir force in a superconducting cavity due to a small
change of temperature.

Considerable change of the dielectric properties occurs
in semiconductor materials that undergo a dielectric-metal
phase transition with an increase in temperature.
It was proposed \cite{80} to measure the change of the
Casimir force acting between an Au-coated sphere and
a vanadium dioxide (VO${}_2$) film deposited on a sapphire
substrate, where the VO${}_2$ undergoes a phase transition.
At $T=68\,{}^{\circ}$C thin VO${}_2$ films undergo an
abrupt transition from a dielectric-type semiconductor to
a metallic phase. During this phase transition the
resistivity of the film decreases by a factor $10^{4}$
from 10 to $10^{-3}\,\Omega\,$cm. The increase in temperature
necessary for the phase transition can be induced by laser
light. Thus, a setup similar to the one described in
Sec.~3.2 can be used. Preliminary theoretical results using
the optical data for VO${}_2$ films show \cite{80,81} that
the difference Casimir force between a Au sphere and VO${}_2$
film after and before the phase transition is sensitive to the
model of conductivity properties used in computations.
Thus, this experiment may help in the resolution of the
problems discussed in Secs.~2 and 3.

Another type of
dielectric-metal phase transitions leading to the change of
dielectric properties is a transition from amorphous to
crystalline state in AgInSbTe.
In Ref.~\cite{82} the gradient of the
Casimir force between a Au coated sphere of $20.2\,\mu$m
diameter and $1\,\mu$m thick AgInSbTe films deposited onto
Al-coated Si substrates was measured using an atomic force
microscope in dynamic mode. Half of the films were amorphous
and the other half were annealed to the
metallic crystalline phase.
It was shown that differences in the gradient of the Casimir
force between a sphere and an amorphous film on the one hand
and between a sphere and a crystalline film on the other
hand achieve 20\% at separations of about 100\,nm.
The obtained experimental results were compared with
computations using the Lifshitz theory and significant
deviations were found. Taking into account that according
to Ref.~\cite{82} the role of surface roughness was negligible at
separations above 70\,nm, these deviations might be explained
by the influence of localized charges in the amorphous phase
and by the use of the Drude model in the extrapolation of
the optical data in the crystalline phase where AgInSbTe
exhitits metallic conductivity (see Secs.~2 and 3).

\subsection{Prospects of using magnetic materials}

The possibility to modify the magnitude of the Casimir force and
even to change attraction for repulsion has long been discussed in
the literature. Specifically, in Ref.~\cite{83} the Casimir
pressure between two magnetodielectric plates was investigated in
the approximation of frequency-independent $\varepsilon$ and $\mu$.
For sufficiently large $\mu$ repulsive forces were found.
It was commented in the literature, however, that for real
magnetic materials $\mu$ is nearly equal to unity in the range of
frequencies, which gives major contribution to the Casimir force
\cite{84}.  As a result, the magnitude of $\mu$ always remains
far away from the values needed to get the Casimir repulsion.
In fact the largest magnitude of $\mu(i\xi)$ is achieved at
$\xi=0$. For diamagnets \cite{85}
$|\mu(0)-1|\sim 10^{-5}$ always. Thus, these materials cannot be used
for modifying the magnitude of the Casimir force. For paramagnets,
with the single exception of ferromagnets, the deviation of
$\mu(0)$ from unity is also negligibly small at any
temperature \cite{85}.

The subset of paramagnetic materials called {\it ferromagnets}
requires special attention when the aim is to modify the Casimir
force. For such materials $\mu(0)\gg 1$ at temperatures lower than
the so-called {\it Curie} temperature  $T_c$ \cite{85}.
The values of $\mu(i\xi)$ for ferromagnets, however, quickly
decrease with the increase of frequency. Thus, for ferromagnetic
metals $\mu(i\xi)$ drops toward unity at frequencies of order
$10^{5}\,$Hz and for ferromagnetic dielectrics at frequencies of order
$10^{9}\,$Hz (note that the first Matsubara frequency $\xi_1$ at
$T=300\,$K is of order $10^{14}\,$Hz). Because of this, in all
computations using Eq.~(\ref{eq1}) at, say, $T>0.3\,$K one can
put $\mu(i\xi_l)=1$ at all $l\geq 1$ and include ferromagnetic
properties only in the zero-frequency term with $l=0$ \cite{86,87}.

Recently the formalism of functional determinants discussed in
Sec.~2 was applied to calculate the Casimir force acting between
an ideal metal cylinder far away from a magnetodielectric plate
\cite{88}. Calculations were performed at $T=0$. It was argued
that for a cylinder remote sifficiently far from the plate, it is
justified to use the static values $\varepsilon(0)$ and $\mu(0)$
for the dielectric permittivity and static permeability of the plate
material. Under these assumptions, both attractive and repulsive
Casimir force was obtained depending on the values of
$\varepsilon(0)$ and $\mu(0)$. It should be remarked, however,
that in order for magnetic properties to influence the
Casimir force, the characteristic frequency
$\omega_c=c/(2a)$ must be less than $10^{9}\,$Hz. Hence it
follows that cylinder-plate separation must be larger than 15\,cm.
At such large separations any discussion concerning the sign of
the Casimir force seems problematic, because this force is equal
to zero for all practical purposes and is considerably
dominated by Newtonian gravity. As was shown in Ref.~\cite{89},
the calculational results for the Casimir force between
magnetodielectric materials with frequency-dependent $\mu$
depend crucially on whether the zero-temperature or thermal
Lifshitz formula is used.

The investigation of thermal Casimir interaction between two
magnetodielectric plates with account of real material properties
was performed in Refs.~\cite{86,87}. As is clear from the
foregoing, only ferromagnets may show noticeable effect on the
Casimir force. With respect to the Casimir effect, however, it is
not reasonable to consider parallel plates made of so-called
{\it hard} ferromagnetic materials which possess spontaneous
magnetization. The point is that the magnetic interaction
between such plates far exceeds any conceivable Casimir force.
Ferromagnetic materials discussed below are what is referred to
as {\it soft} ferromagnetic materials,  which do not  possess spontaneous
magnetization. It is well known also \cite{85} that the magnetic
permeability of ferromagnets depends on the applied magnetic field
(the so-called {\it hysteresis}). Keeping in mind that no external
magnetic field is applied to the Casimir plates and that the mean
value of the fluctuating magnetic field is equal to zero, one
should consider what is often referred to as  {\it initial}
permeability, i.e., $\mu(\mbox{\boldmath$H$}=0)$.

Now we present a few computational results illustrating the impact
of ferromagnetic materials on the magnitude and sign of the
Casimir force. Computations of the Casimir pressure were performed
using Eq.~(\ref{eq1}) at room temperature with ferromagnetic
properties included in the zero-frequency term $l=0$.
The dielectric properties of ferromagnetic and nonmagnetic
metallic plates are described either by the Drude model (\ref{eq13})
or by the plasma model (\ref{eq14}) and the obtained results are
compared. Keeping in mind that ferromagnetic properties contribute
only through the zero-frequency term, it is reasonable to consider
not too short separations where the relative contribution of this
term is more pronounced. Below we consider separation distances
from 0.5 to $6\,\mu$m, where the role of core electrons in the
dielectric response can be neglected.

First we consider the Casimir pressure between two plates made
of the ferromagnetic metal Co with $\mu_{\rm Co}^{(n)}(0)=70$
\cite{90}. The Drude parameters of Co are the following
\cite{91}: $\omega_{p,{\rm Co}}^{(n)}=3.97\,$eV and
$\gamma_{\rm Co}^{(n)}=0.036\,$eV. The computational results
are presented as a ratio to the zero-temperature Casimir
pressure between two nonmagnetic plates made of ideal metal
at zero temperature
\begin{equation}
P_{0}(a)=-\frac{\pi^2}{240}\,\frac{\hbar c}{a^4}.
\label{eq25}
\end{equation}
\noindent
In Fig.~8, the solid and dashed lines show the values of
$P_{\rm Co}/P_0$ as functions of separation computed using the
dielectric permittivity (a) of the Drude model and (b) of the
plasma model with account of magnetic properties and with
magnetic properties disregarded, respectively.
\begin{figure}[t]
\begin{center}
\vspace*{-7.5cm}
\hspace*{-4.cm}
\includegraphics[width=22cm]{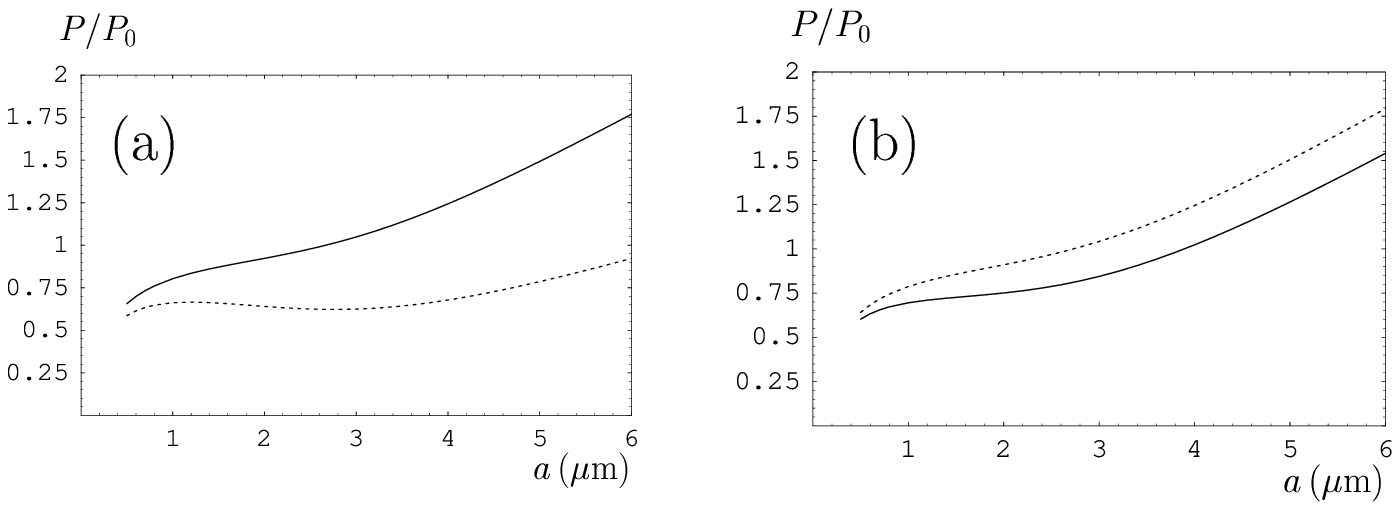}
\vspace*{-19.3cm}
\end{center}\caption{The relative Casimir
 pressures as functions of separation
in the configuration of two parallel Co plates with account of
magnetic properties (solid lines) and with magnetic
properties disregarded (dashed lines). Computations are
performed at $T=300\,$K using  (a) the Drude model
and (b) the plasma model for the dielectric permittivity.
}
\end{figure}
Quantitatively, the role of magnetic properties can be
characterized by the ratio
\begin{equation}
\eta_{P,{\rm Co}}(a,T)=
\frac{P_{\rm Co}^{\rm solid}(a,T)-
P_{\rm Co}^{\rm dashed}(a,T)}{P_{\rm Co}^{\rm dashed}(a,T)}.
\label{eq26}
\end{equation}
\noindent
With the increase in separation from 0.5 to $2\,\mu$m and then
to $6\,\mu$m, $\eta_{P,{\rm Co}}^{D}$ computed using the Drude
model varies from 12\% to 44\% and to 92\%, respectively.
Thus, when the Drude model is used to describe the dielectric
properties of a ferromagnetic metal, the magnetic properties
markedly (up to almost two times at large separations)
increase the magnitude of the Casimir pressure \cite{86}.
As can be seen in Fig.~8(b), for ferromagnetic metal described
by the plasma model the impact of magnetic properties on the
Casimir pressure is not so pronounced. Quantitatively,
from Fig.~8(b) it follows that at separation 0.5, 2, and
$6\,\mu$m $\eta_{P,{\rm Co}}^{p}$ varies from --6\% to --17\%,
and to --14\%, respectively. Mention should be made that if
the plasma model is used, the inclusion of magnetic
properties decreases the magnitude of the Casimir pressure.
Similar results were also obtained for Fe \cite{86}.
In both cases the inclusion of the magnetic properties
does not change the sign of the Casimir force leaving it
attractive.

Very prospective materials for the investigation of the impact of
magnetic properties on the Casimir force are ferromagnetic
dielectrics. These materials, while displaying physical
properties characteristic for dielectrics, demonstrate ferromagnetic
behavior under the influence of an external magnetic field \cite{92}.
Many ferromagnetic dielectrics are, for instance, composite
materials obtained on the basis of a matrix of a polymer compound with
inclusion of nanoparticles of ferromagnetic metals. As was shown
in Ref.~\cite{86}, the Casimir force between two parallel plates
made of ferromagnetic dielectric remains attractive, while the
influence of magnetic properties on the Casimir pressure is much
stronger than for Co plates considered above.

Here, we consider the configuration of one plate made
of ferromagnetic dielectric ($n=1$) and the other plate made of
a nonmagnetic metal Au ($n=2$). As is ahown below, in this
configuration the results for the Casimir pressure are
drastically different depending on whether the Drude or the
plasma model is used for the description of Au. In the case when
the Drude model is used, from Eqs.~(\ref{eq3}) and (\ref{eq13})
it follows that $r_{\rm TE}^{(2)}(0,k_{\bot})=0$.
Keeping in mind that at zero frequency magnetic properties do not
influence the TM reflection coefficient, we arrive at the
conclusion that if the Drude model is used for the description
of Au there is no impact of magnetic properties on the Casimir
pressure between gold and magnetodielectric plates.
To perform computations when Au plate is described by the plasma
model, one should choose some specific ferromagnetic dielecric.
We consider the composite material on the basis of polystyrene
with the volume fraction of ferromagnetic particles in the
mixture $f=0.25$. The magnetic permeability of such
materials may vary over a wide range \cite{93}. Below we use
$\mu(0)=25$. The dielectric permittivity of polystyrene
$\varepsilon_d(i\xi)$ is presented in the form (\ref{eq9})
with $K=4$. The parameters of oscillators can be found in
Ref.~\cite{32} leading to $\varepsilon_d(0)=2.56$.
As a result, the dielectric permittivity of the used
ferromagnetic dielectric can be presented in the form \cite{95}
\begin{equation}
\varepsilon_{\rm fd}^{(1)}(i\xi)=\varepsilon_d(i\xi)
\left(1+\frac{3f}{1-f}\right).
\label{eq27}
\end{equation}
\noindent
The dielectric permittivity of Au is described by Eq.~(\ref{eq14})
with $n=2$, $\omega_p^{(2)}=9.0\,$eV.

\begin{figure}[t]
\begin{center}
\vspace*{-7.5cm}
\hspace*{-4.2cm}
\includegraphics[width=22cm]{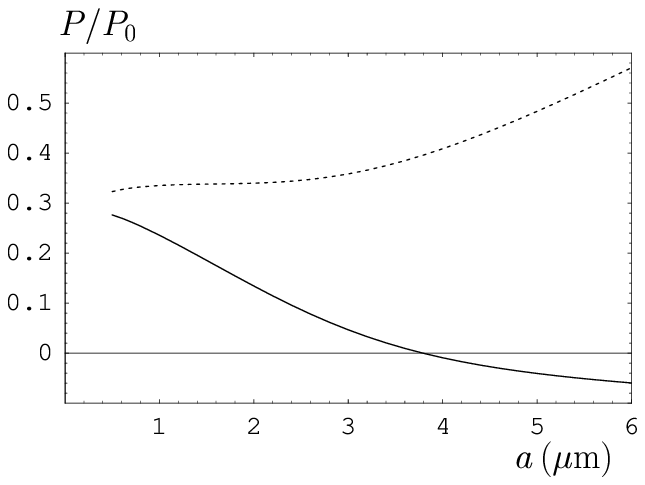}
\vspace*{-19.3cm}
\end{center}\caption{ The relative Casimir
 pressures as functions of separation
 at $T=300\,$K
in the configuration of one plate made of
ferromagnetic dielectric and the other
plate made of Au with account of
magnetic properties (solid line) and with magnetic
properties disregarded (dashed line). Computations are
performed using the plasma model for the dielectric permittivity
of Au.
}
\end{figure}
The computational results for the relative Casimir pressure
$P/P_0$ at $T=300\,$K as a function of separation are presented in
Fig.~9. The solid (dashed) lines show the results obtained with
magnetic properties of ferromagnetic plate included
(disregarded). As is seen in Fig.~9, there is the profound
effect of magnetic properties of ferromagnetic dielectric on
the Casimir pressure. Thus, at separations of 0.5, 2, and
$6\,\mu$m the values of $\eta_{P,{\rm fd-Au}}^{p}$ are equal
to --14\%, --60\%, and --110\%, respectively.
What is more important, the Casimir pressure $P$ (and, thus,
the Casimir force acting between parallel plates through a
vacuum gap) changes sign and becomes positive at separations
$a>3.8\,\mu$m. This means that the Casimir force becomes
repulsive if Au plate is described by the plasma model.
The effect of repulsion between a nonmagnetic metal plate and
a ferromagnetic dielectric plate can be used as an experimental test for
the influence of magnetic properties on the Casimir force and
for the model of dielectric permittivity of a metal plate.

As was mentioned in the beginning of this section, at the Curie
temperature $T_C$ specific for each material, ferromagnets
undergo a phase transition \cite{96}. At higher temperature
they become paramagnets with negligibly small magnetic
properties with respect to the Casimir force. The Curie
temperature varies over a wide temperature region (for Co,
for instance, it is equal to 1388\,K). Keeping in mind
to investigate the behavior of the Casimir pressure under the
magnetic phase transition, it is preferable to deal with two
plates made of ferromagnetic metal possessing the Curie
temperature close to room temperature, such as Gd
($T_C$ is of about 290\,K). Dielectric properties of Gd
are described by either the Drude or the plasma model with
the parameters \cite{97} $\omega_{p,{\rm Gd}}=9.1\,$eV and
$\gamma_{\rm Gd}=0.58\,$eV. The dependence of static
magnetic permeability of Gd as a function of temperature
in the vicinity of Curie temperature in modeled in Fig.~10(a)
using the data of Ref.~\cite{98}.

\begin{figure}[t]
\begin{center}
\vspace*{-7.5cm}
\hspace*{-4.2cm}
\includegraphics[width=22cm]{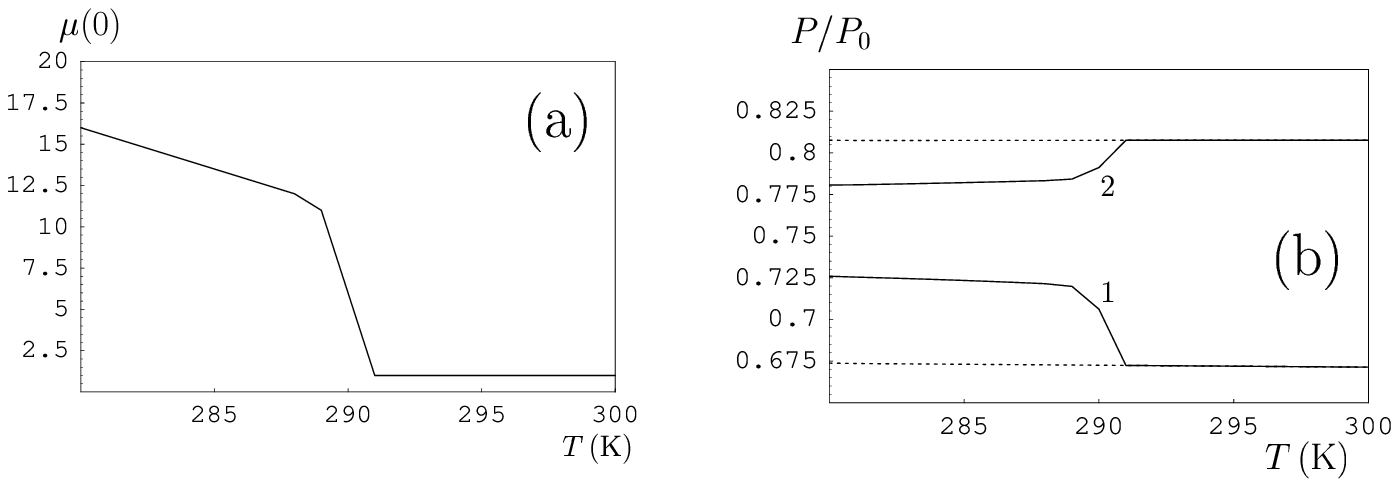}
\vspace*{-19.3cm}
\end{center}\caption{(a) The static magnetic permeability of Gd
in the magnetic phase transition as a function of temperature.
(b) The relative Casimir  pressures as functions of temperature
in the configuration of two parallel Gd plates at the separation
$a=0.5\,\mu$m. The solid and dashed lines take into account and
disregard the magnetic properties, respectively. The pairs
of lines marked 1 and 2 indicate respective computational
results obtained using the Drude and plasma models.
}
\end{figure}
Computations of the relative Casimir pressure $P/P_0$ in the
configuration of two Gd plates were performed at $a=500\,$nm
as a function of temperature using Eq.~(\ref{eq1}).
The computational results are presented in Fig.~10(b), where
the pairs of solid and dashed lines marked 1 and 2 indicate the
results computed using the Drude and plasma models,
respectively. As before, the solid lines take into account
and the dashed lines disregard the magnetic properties
of Gd.  As can be seen in Fig.~10(b), at $T>T_C$ the magnetic
properties do not influence the Casimir pressure, whereas
the Drude and plasma model approaches lead to results
differing for about --19.5\%.
At $T<T_C$ the magnetic properties influence the Casimir
pressure. Quantitatively, at $T=280\,$K the relative
influence of magnetic properties on the Casimir pressure is
$\eta_{P,{\rm Gd}}^{D}=7.4$\% for the Drude model and
$\eta_{P,{\rm Gd}}^{p}=-3.3$\% for the plasma  model.
With account of magnetic properties, the relative difference
between the predictions of the Drude and plasma model
approaches for the Casimir pressure is approximately equal
to --7\%. Thus, the magnetic phase transition provides
additional opportunities for the invistigation of the
impact of magnetic properties on the Casimir force and for
the selection between different models of dielectric
permittivity.

To conclude this section, we note that possible influence
of magnetic properties of ferromagnets on the Casimir force
may be considered as somewhat analogous to the influence
of real drift current of conduction electrons discussed in
Sec.~2. If it is assumed that the fluctuating electromagnetic
field can initiate such a current, we arrive at the Drude
model approach to the thermal Casimir force which is
considered as most natural by some authors \cite{38,39,40}.
This approach, however, was found to be in contradiction with
the results of several experiments \cite{8,9,10,11,12,13,14,15}.
In this connection the problem arises whether the fluctuating
electromagnetic field can lead to magnetic effects in
ferromagnets which remains to be solved.

\section{Lateral Casimir force at the nanoscale}

The most universally known {\it normal} Casimir force acts in
the direction perpendicular to the interacting surfaces.
However, when the material surfaces of the interacting
bodies are anisotropic or they are positioned asymmetrically,
a lateral Casimir force may exist, which acts tangential to
the surface \cite{16}. Experimentally the lateral Casimir
force was first measured \cite{17,18} in the configuration
of a sinusoidally corrugated Au-coated sphere above a
sinusoidally corrugated Au-coated plate with equal corrugation
periods. The measurement data were compared with theoretical
predictions using the PFA. For two parallel ideal metal plates
covered with uniaxial sinusoidal corrugations of equal periods
the lateral Casimir force was calculated using the exact
theory in the second perturbation order with respect to relative
corrugation amplitude \cite{99}. For sinusoidally corrugated
plates made of real metals described by the plasma model,
the lateral Casimir force was calculated in Refs.~\cite{100,101}
also in the second-order perturbation theory.
Experimentally, the lateral Casimir force becomes measurable
only for sufficiently large amplitudes of corrugations and at
sufficiently short separations. Because of this, calculational
methods restricted by the second-order perturbation theory
are usually not sufficient for the comparison with the
measurement data \cite{102}.

In the first measurement of the lateral Casimir force \cite{17,18}
the corrugation amplitudes were $A_1=59\,$nm on the plate and
$A_2=8\,$nm on the sphere. The period of corrugations was equal
to $\Lambda=1200\,$nm, i.e., much larger than the separation between
the sphere and the plate. The lateral Casimir force with an
amplitude of $3.2\times 10^{-13}\,$N at the shortest separation
was found to sinusoidally oscillate as a function of the phase
shift $\varphi$ between the corrugations. It was also predicted
theoretically, but not observed experimentally due to the use
of insufficiently large corrugation amplitudes, that the
lateral Casimir force is in fact asymmetric, i.e., deviates
from purely sinusoidal harmonic dependence on the phase shift
(this phenomenon is absent when computations are performed
up to the second perturbation order).

In the recently performed new measurements of the lateral
Casimir force between the Au-coated sphere and plate covered
with uniaxial sinusoidal corrugations, the corrugation
amplitudes were $A_1=85.4\,$nm on the plate and
$A_2=13.7\,$nm on the sphere ($\tilde{A}_2=25.5\,$nm
in the second set of measurements).
The corrugations had an average period $\Lambda=574.7\,$nm,
i.e., more than two times smaller than in Refs.~\cite{17,18}
in order to achieve the regime where the PFA becomes
inapplicable to the calculation of the lateral Casimir force
[see Fig.~11(a,b) where the corrugated surfaces of the plate
and of the sphere used in the first set of measurements,
respectively, are shown]. To obtain corrugations on the
sphere with exactly the same period, as on the plate,
a homogeneous sphere was sandwiched against the plate and
pressed by means of a piezo. In such a manner sinusoidal
corrugations from the plate were imprinted on the sphere.
This was done preserving the sphericity of a sphere
(see Fig.~12, where the lighter tone shows higher points
on the surface of the sphere used in the first set of
measurements).
\begin{figure}[b]
\begin{center}
\vspace*{-1.3cm}
\hspace*{-3.8cm}
\includegraphics[width=21.5cm]{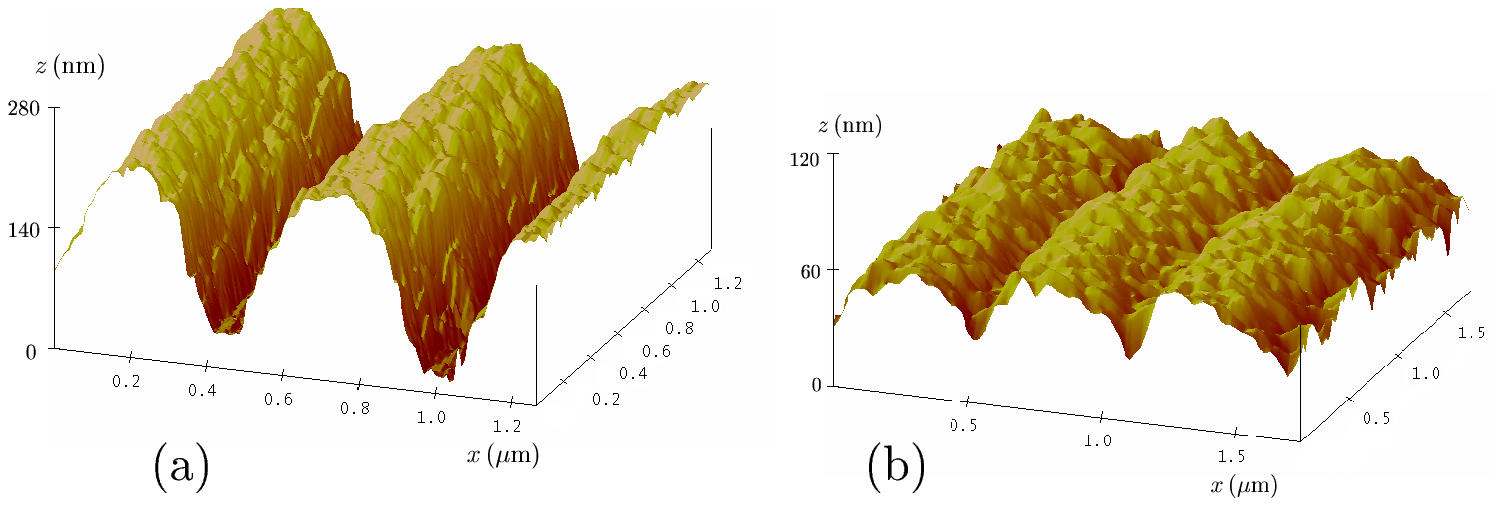}
\vspace*{-24.5cm}
\end{center}\caption{An AFM scan
(a) of the plate surface and (b) of the  surface of the sphere
showing the sinusoidal
corrugations covered with stochastic roughness.
}
\end{figure}

\begin{figure}[t]
\begin{center}
\vspace*{-7.5cm}
\hspace*{-.2cm}
\includegraphics[width=16cm]{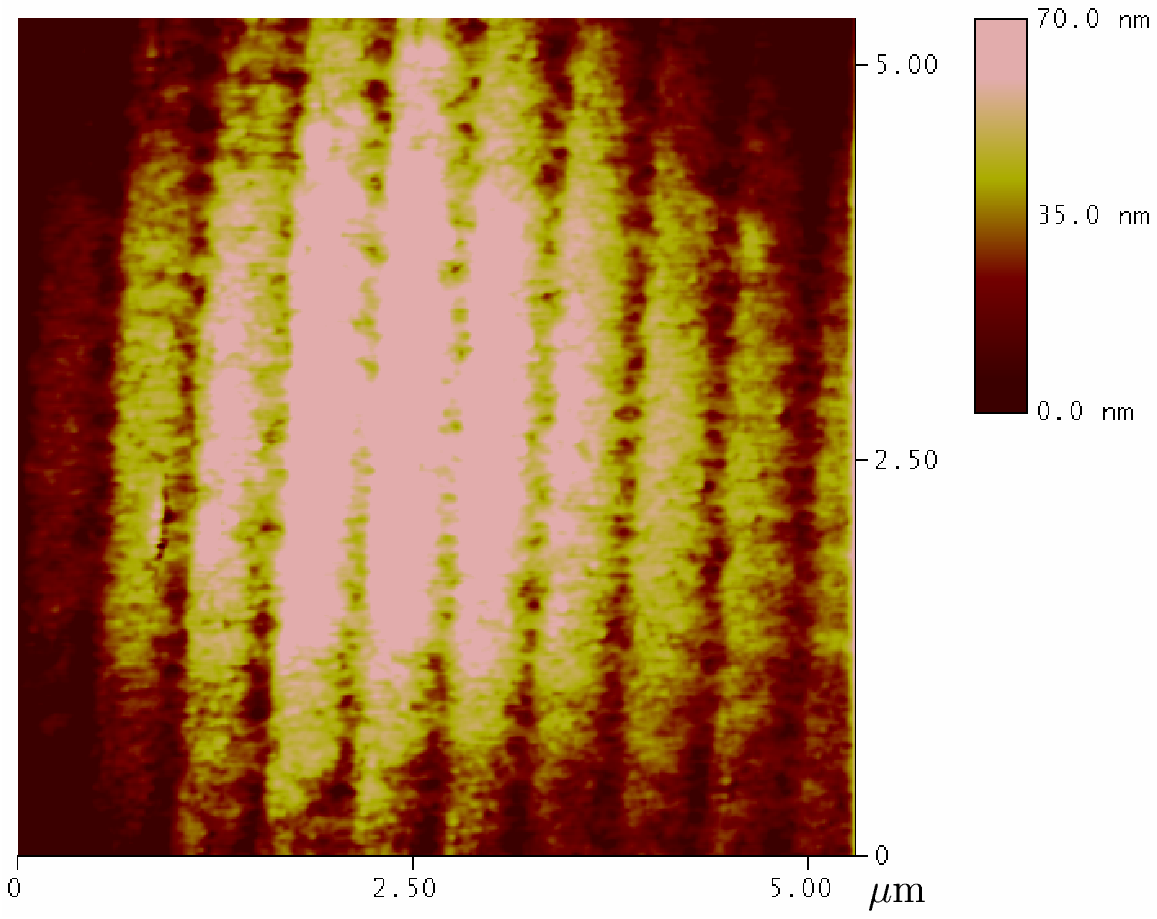}
\vspace*{-8.3cm}
\end{center}\caption{ The imprinted
corrugations on the sphere.
The lighter area shows higher points and hence
demonstrates the sphericity of the imprinted surface.
}
\end{figure}

Taking into account the sphericity of the second body and
sinusoidal corrugations on both bodies in the framework
of the PFA and using the Lifshitz formula (\ref{eq1}),
one arrives to the following approximate expression for
the lateral Casimir force \cite{20}
\begin{equation}
F_{\rm lat}(a,T,\varphi)=
\frac{\pi k_BTRA_1A_2}{2a^3\Lambda\beta}\,\sin\varphi\,
\sum_{n=1}^{\infty}
\int_{\zeta_l}^{\infty}y\,dy\,e^{-ny}\,I_1(n\beta y)
\left[r_{\rm TM}^{2n}(i\zeta_l,y)+
r_{\rm TE}^{2n}(i\zeta_l,y)\right].
\label{eq28}
\end{equation}
\noindent
Here,
\begin{equation}
\beta\equiv\beta(a,\varphi)=\frac{1}{a}
(A_1^2+A_2^2-2A_1A_2\cos\varphi)^{1/2},
\label{eq29}
\end{equation}
\noindent
$I_1(z)$ is the Bessel function of an imaginary argument, and
the following dimensionless variables are used:
\begin{equation}
y=2aq_l,\qquad \zeta_l=\frac{2a\xi_l}{c}.
\label{eq30}
\end{equation}
\noindent
Remembering that Eq.~(\ref{eq28}) uses the PFA for the description
of sinusoidal corrugations, it should lead to sufficiently exact
results under a condition $2\pi a\ll\Lambda$ \cite{6,7a}.
As to the applicability of the PFA to sphere-plate geometry,
this is garanteed with large safety margin because
$a\ll R=97\,\mu$m.

Measurements of the lateral Casimir force in Refs.~\cite{19,20}
were performed at separations between the mean levels of
corrugations of about 100--200\,nm. Because of this, the
application condition of the PFA for the description of corrugations
was not satisfied. Therefore, some corrections to Eq.~(\ref{eq28})
due to diffraction-type effects were expected to arise.
To take the diffraction-type effects into account, the exact
description of corrugations using the Rayleigh scattering
theory was developed  \cite{19,20} while preserving the use
of the PFA for the description of sphere-plate geometry, where
it is sufficiently accurate. As a result, the following
``exact" expression for the lateral Casimir force was
obtained \cite{19,20}
\begin{eqnarray}
F_{\rm lat}^{\rm exact}(a,T,\varphi)&=&\frac{4k_BTR}{\Lambda}
\int_{a}^{\infty}{\!\!\!}dz\sum_{l=0}^{\infty}{\vphantom{\sum}}^{\prime}
\frac{\partial}{\partial\varphi}\int_{0}^{\infty{\!\!\!}}dk_y
\int_{0}^{\pi/\Lambda}{\!\!}dk_x
\label{eq31} \\
&\times&
\ln\,\mbox{det}\left[I-
R_{\rm bot}^{(1)}(\mbox{\boldmath$k$}_{\bot},i\xi_l)
R_{\rm top}^{(2)}(\mbox{\boldmath$k$}_{\bot},i\xi_l,
z+A_1+A_2,\varphi)\right].
\nonumber
\end{eqnarray}
\noindent
Here, $R_{\rm bot}^{(1)}$ ($R_{\rm top}^{(2)}$) is the reflection
matrix of the downward (upward) moving waves on the bottom (top)
grating under a condition that the top (bottom) grating is
removed, and $I$ is an infinite dimensional unit matrix.
In both Eqs.~(\ref{eq28}) and (\ref{eq31}) the dielectric
properties of Au were taken into account by means of the
generalized plasma-like model (\ref{eq21}).

\begin{figure}[t]
\begin{center}
\vspace*{-1.3cm}
\hspace*{-3.2cm}
\includegraphics[width=22.5cm]{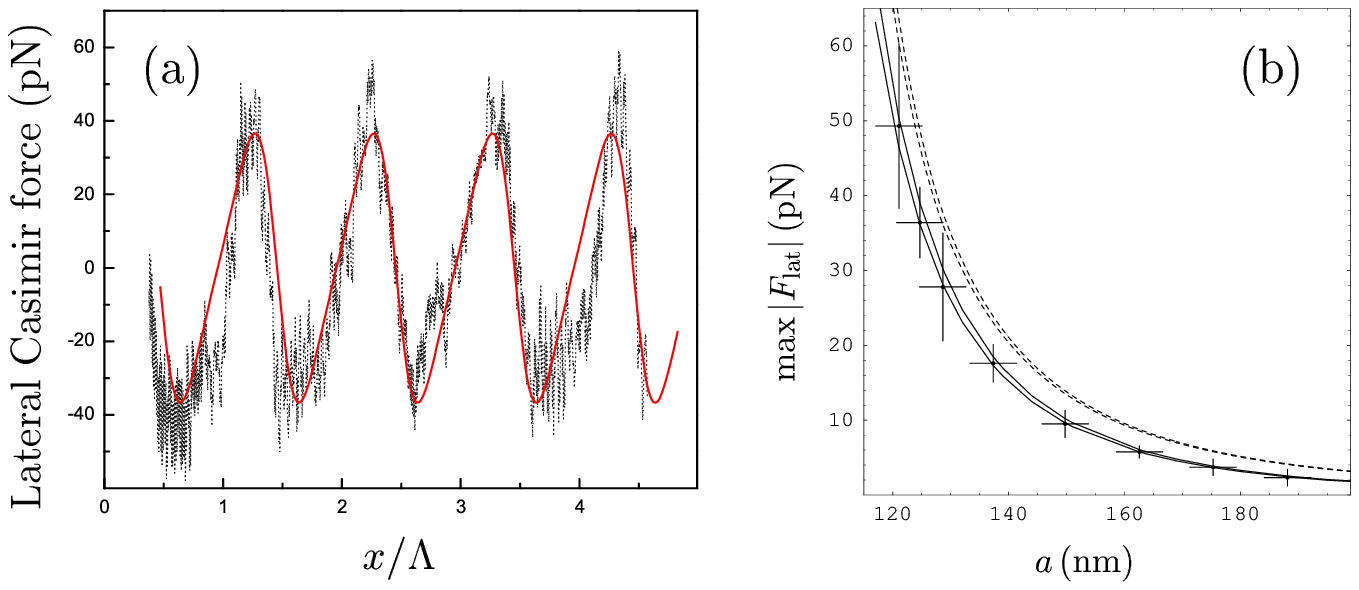}
\vspace*{-25.cm}
\end{center}\caption{(a) The
experiment (dots) and the exact theory (solid
line) for the lateral Casimir force versus the
lateral displacement normalized for the
corrugation period at the separation 124.7\,nm.
No fitting parameters are used.
(b) The
experimental data (crosses) and theoretical values
computed using the exact theory
(the band between the solid lines) and using the proximity force
approximation (the band between the dashed lines)
for the maximum magnitudes of the
lateral Casimir force versus  separation.
No fitting parameters are used.
}
\end{figure}
In Fig.~13(a) we show the measured data in the first set of
measurements (dots) and computed using the exact
theory of Eq.~(\ref{eq31}) lateral Casimir force (solid line)
over four corrugation periods versus the normalized lateral
displacement $x/\Lambda$ between the corrugated surfaces at
a separation $a=124.7\,$nm (similar figures related to larger
separations and to the second set of measurements can be
found in Ref.~\cite{20}). As can be seen in Fig.~13(a),
the lateral Casimir force is a periodic function of the phase
shift. The important characteristic feature of the periodic
curve shown in the figure is that it is {\it asymmetric},
i.e., the dependence of $F_{\rm lat}$ on $x/\Lambda$ is not
strictly sinusoidal (note that the sinusoidal dependence
of the lateral force on $\varphi=2\pi x/\Lambda$ holds
only if the calculations are restricted to the second order
in corrugation amplitudes $\sim A_1A_2$).
The true asymmetry of the measured lateral force is obvious even
without the theory line. For example, in Fig.~13(a) the average
shift of each maximum point from the midpoint of two adjacent
minima is $(0.12\pm 0.01)\Lambda$. We emphasize that the
theoretical solid line was obtained with no fitting parameters.
As can be seen in Fig.~13(a), the exact theory of Eq.~(\ref{eq31})
is in a very good agreement with the experimental data shown as
dots.

The experimental data for the $\max|F_{\rm lat}|$ versus
separation $a$ are shown in Fig.~13(b) as crosses.
The arms of crosses indicate the total experimental error
determined at a 95\% confidence level for each experimental
point individually. The exact computational results for
the $\max|F_{\rm lat}|$ are shown in Fig.~13(b) as a band
between the two solid lines versus separation. The width of this
band takes into account the computational errors and the
correction to Eq.~(\ref{eq31}) due to surface roughness.
As is seen in Fig.~13(b), the exact theory is in a very good
agreement with the measurement data. The computational results for
the $\max|F_{\rm lat}|$ at different separations obtained using
the PFA [Eq.~(\ref{eq28})] are shown in Fig.~13(b) as a band
between two dashed lines. The width of this band takes into
account the role of surface roughness. As can be seen in
Fig.~13(b), the experimental data indicated as crosses are
inconsistent with the prediction of the PFA approach, thus
revealing the role of diffraction-type effects for
corrugations of relatively small period used in this
experiment.

Besides fundamental interest in the phenomenon of lateral
Casimir force as one more manifestation of the zero-point
oscillations of electromagnetic field, it is pertinent to note
that this version of the Casimir force gives the possibility
to actuate lateral translations of corrugated surfaces in
micromachines. The lateral Casimir force might be used to
solve the tribological problems plaguing the microdevice
industry. Specifically, it was proposed to use this effect
for the frictionless transmission of lateral motion by
means of a nanoscale rack and pinion without intermeshing
cogs \cite{103,104,104a} or in a ratchet with asymmetric
corrugations driven by the Casimir force \cite{105}.
Such devices could be used for transfering motion
between two microelectromechanical systems without
mechanical contact. The experimental and theoretical
results on the lateral Casimir force discussed above
bring us closer to the realization of micromachines
driven by the zero-point fluctuations of quantum
vacuum.

\section{Conclusions and discussion}

In the foregoing, we have listed several experiments and
theoretical advances in the Casimir effect which has come to the
center of public attention very recently.
The Casimir effect deals with quantum vacuum, which is the
most fundamental and puzzling type of physical reality.
This is a possible reason why the research area in this field is
so wide. On the one hand, it is closely connected with
foundations of quantum statistical physics and even calls for
reconsideration of some of its basic concepts. On the other
hand, the Casimir effect quickly turns into a practical tool
of nanotechnology which is expected to find numerous
applications in the next generation of microdevices.

As was discussed in Sec.~2, the fundamental Lifshitz theory
of the van der Waals and Casimir forces comes into conflict
with basics of thermodynamics when the relaxation
properties of free charge carriers are taken into account.
In Sec.~3, the three landmark recently performed experiments
are described demonstrating that in a literal sense, the
Lifshitz theory is in a deep contradiction with the
measurement data. Only if we accept the phenomenological
prescription that for dielectrics charge carriers are
disregarded and for metals they are described by the plasma
model, the theoretical predictions of the Lifshitz theory come
in agreement with the data. This situation cannot be
considered as satisfactory. In fact we deal with some kind
of a paradox. The mathematical formalism of the Lifshitz
theory during the last few years has undergone far reaching
generalization discussed in Secs.~2 and 5. It is presently
possible to write the Lifshitz-type formulas for compact
objects of arbitrary shape in terms of reflection amplitudes
of electromagnetic oscillations on their surfaces.
Based on this it was even believed that at the present time
computation of the Casimir force between any bodies is only
a problem of sufficient computer facilities.
As was argued in Secs.~2 and 3, the problem is, however,
much deeper. In fact the reflection amplitudes suggested
by our knowledge of physical processes with real (not
fluctuating) electromagnetic fields lead to
computational results in contradiction with
the experimental data
making powerful mathematical formalism at least insufficient.
This is a fundamental problem to be solved in the future.

{}From the point of view of applications in nanotechnology,
large progress on the control the magnitude of the Casimir
force was achieved. This was done by means of a modification
of dielectric properties by using different materials and
different influential factors such as irradiation with
laser light. In several experiments considered in Sec.~4.1 it was
possible to modify the magnitude of the Casimir force from a
few percent to about 50\%. There were several proposals
on how to achieve the Casimir repulsion through the vacuum gap,
but to the present day the Casimir repulsion was realized only
in the familiar three-layer system. One more opportunity to
control the magnitude of the Casimir force is suggested by phase
transitions. In Sec.~4.1 we have discussed the phase
transition from dielectric to the metallic state. A few tens
of percent
variation in the magnitude of the Casimir force is achievable
under a phase transition of the plate material.

Quick progress in the application of the Casimir force in
nanotechnology might be caused by the use of ferromagnetic
materials. For real material properties this subject was
investigated only recently (see Sec.~4.2).
What is most striking, according to the Lifshitz theory the
Casimir interaction of a nonmagnetic
metallic  plate described by the
plasma model with a ferromagnetic dielectric plate through
the vacuum gap becomes repulsive under some conditions.
This effect, if exists in nature, opens up the way for
the developments of various microdevices. We should notice,
however, that the influence of magnetic properties on the
Casimir interaction is not yet confirmed experimentally.
The confirmation or exclusion of such an influence is
expected in the near future.

One more opportunity for applications in nanotechnology
is suggested by the lateral Casimir force discussed in
Sec.~5. Together with the normal Casimir force, it gives
the possibility to actuate both normal and lateral
translations of corrugated surfaces in micromachines
by means of the electromagnetic zero-point oscillations.
With respect to tribological problems, the lateral
Casimir force allows frictionless transition of lateral
motion. All the above allows to conclude that the
Casimir effect promises important breakthroughs in both
fundamental physics and its applications in nanotechnology.

\ack{
The author is grateful to the Organizing Committee of ICSF2010
for the invitation to present this talk, kind hospitality
during the Conference and financial support. CNPq (Brazil),
process 310039/2010--0 is also acknowledged for partial financial
support. I am grateful to all co-authors of joint papers for
helpful discussions and collaboration.
}

\end{document}